\documentclass[preprint2]{aastex}

\usepackage{natbib}

\slugcomment{Not to appear in Nonlearned J., 45.}

\shorttitle{Chandra detected SNRs in nearby galaxies}
\shortauthors{Leonidaki et al.}

\begin{document}

\title{A multiwavelength study of Supernova Remnants in six nearby galaxies.\\
I: Detection of new X--ray selected Supernova Remnants with Chandra}

\author{I.Leonidaki\altaffilmark{1,3}, A.Zezas\altaffilmark{2,4,5} and P.Boumis\altaffilmark{1}}
\affil{National Observatory of Athens, Institute of Astronomy and Astrophysics}
\affil{Harvard-Smithsonian Center for Astrophysics}
\affil{Astronomical Laboratory, Physics Department, University of Patras}
\affil{Physics Department, University of Crete}
\affil{IESL / Foundation for Research and Technology - Hellas}

\altaffiltext{1}{I.Metaxa and V.Pavlou, Lofos Koufou, Penteli, 15236, Athens, Greece}
\altaffiltext{2}{60 Garden Street, Cambridge, MA 02138, USA}
\altaffiltext{3}{26500, Rio, Patra , Greece}
\altaffiltext{4}{P.O Box 2208, GR - 710 03, Heraklion, Crete, Greece}
\altaffiltext{5}{P.O Box 1527, GR - 711 10, Heraklion, Crete, Greece}

\begin{abstract}
We present results from a study of the Supernova Remnant (SNR) population in a sample of six nearby galaxies (NGC 2403, NGC 3077, NGC 4214, NGC 4449, NGC 4395 and NGC 5204) based on {\it Chandra} archival data. We have detected 244 discrete X-ray sources down to a limiting flux of 10$^{-15}$ erg s$^{-1}$. We identify 37 X-ray selected thermal SNRs based on their X--ray colors or spectra, 30 of which are new discoveries. In many cases the X--ray classification is confirmed based on counterparts with SNRs identified in other wavelengths. Three of the galaxies in our sample (NGC 4214, NGC 4395 and NGC 5204) are studied for the first time, resulting in the discovery of 13 thermal SNRs. We discuss the properties (luminosity, temperature, density) of the X-ray detected SNRs in the galaxies of our sample in order to address their dependence on their environment. We find that X--ray selected SNRs in irregular galaxies appear to be more luminous than those in spirals. We attribute this to the lower metallicities and therefore more massive progenitor stars of irregular galaxies or the higher local densities of the ISM. We also discuss the X--ray selected SNR populations in the context of the Star Formation Rate of their host galaxies. A comparison of the numbers of observed luminous X--ray selected SNRs with those expected based on the luminosity functions of X--ray SNRs in the MCs and M33 suggest different luminosity distributions between the SNRs in spiral and irregular galaxies with the latter tending to have flatter distributions.
\end{abstract}


\section{Introduction}

Studies of the populations and evolution of Supernova Remnants (SNRs)
can provide information on the interplay between massive--star
formation and the local interstellar medium (ISM; e.g. \citealt{Chu95,Bykov05}). SNRs provide a significant fraction of
the mechanical energy that heats, shapes and chemically enriches the
ISM. Therefore, SNRs can be used to investigate global properties of
the galaxy's ISM as well as their local environment 
\citep{BL04}. Furthermore, they can be used as proxies to measure the
formation of massive stars and can give information on star formation
rate and stellar evolution \citep{CY90}.

Detecting large samples of SNRs in more than one wavebands can provide information
about the different processes which take place during their
evolution. It is expected that radiation of newly formed SNRs is
dominated by the hot material behind the shock--wave producing thermal
X--rays (thermal SNRs). One special category are plerion--type SNRs
where non--thermal processes are the dominant X--ray emission mechanism
(e.g. \citealt{Safi01, Asaoka90}). Optical filaments are a sign of
older SNRs since they form in the cooling regions behind the shock
(e.g. \citealt{CS, Stupar09}). Emission in radio wavelength is radiated from the vicinity
of the shock as well as from the cooling filaments and is easily
detectable throughout the life of the remnant (e.g. \citealt{Dickel99, CS}). In cases
where a source is detected in two or more wavebands (e.g. X-rays and
optical), it is an indication of material at a wide range of
temperatures and the existence of high energy electrons
\citep{CS}. Because of these selection effects, only 
multiwavelength studies (X-rays, radio, optical, infrared) of large
samples of extragalactic SNRs will give a complete image about their
nature and evolution as well as their correlation with
star--formation.

About 274 SNRs are known to exist in our Galaxy \citep{Green09} and a
large number of them has been studied in detail. While these studies
provide important information on the properties of the individual
objects and SNR physics, studies of the population of SNRs are
hampered by absorption due to the dust of the Galaxy and distance
uncertainties. Therefore, the study of nearby galaxies offers several
advantages: the entire galaxy can be studied with fewer observations,
SNRs are at the same distance and by selecting higher Galactic
latitude, face--on galaxies, internal galactic absorption effects are
minimized.

Previous X-ray observations of nearby galaxies ($\la$ 5
Mpc) with ROSAT (e.g. \citealp{S94, S00, P00, P02, Payne04}) revealed
several X--ray emitting SNRs some of which were also found in other
wavebands. These observations showed that SNRs are an important component of the X-ray source populations in
galaxies \citep{BL97}, especially at luminosities below 10$^{37}$ erg
s$^{-1}$. However, the ROSAT surveys were limited by their low
sensitivity and spatial resolution and confined the detection and study of discrete X--ray sources only within our Galaxy and the Local Group. The much better capabilities of {\it Chandra} in both domains
offer a unique opportunity to detect large populations of SNRs in
nearby galaxies. The excellent angular resolution of {\it Chandra}
($\thicksim$ 0.5$\arcsec$) reveals X-ray sources that could not be
clearly detected otherwise and allows their identification with sources detected
in other wavebands. With typical detection limits of $\thicksim$
10$^{37}$ erg s$^{-1}$ in moderate exposures, extragalactic Cas--A
or Crab--like SNRs can be detected in nearby galaxies. The capability of {\it Chandra} to
detect X-ray emitting SNRs has been demonstrated in
several studies of nearby galaxies; e.g. NGC 1637 --
\citep{Immler03}; NGC 6822 -- \citep{Kong04}; M 31 --
\citep{Kong02}; NGC 2403 -- \citep{SP03}; M 81 --
\citep{Swartz03}.

Even though, about 650, 200 and 50 extragalactic SNRs have been
discovered to date in the optical, radio and X--ray bands
(e.g. \citealp{BL97, MF97, MFBL97}), these numbers come only from a
small set of nearby galaxies. Moreover, there is a large gap between
detection rates in different bands (less than 50 SNRs have been
identified in all three bands) not only due to evolutionary
effects but also because of the different observational sensitivity and
resolution. In the X--ray band in particular there have not been any systematic studies of the SNR populations in nearby galaxies. Most importantly all studies of X--ray emitting SNRs outside the Local Group have been focused on the identification of X--ray counterparts to SNRs detected in other wavebands, rather than searching for new X--ray emitting SNRs.

This paper is the first in a series undertaking a multiwavelength study
of extragalactic SNR populations. We focus on X--ray emitting thermal
SNRs rather than plerions because the latter have very similar X--ray
spectra to X--ray Binaries (XRBs) and therefore they cannot be
identified based on their X--ray properties alone. When possible, we
cross--correlate our results with multiwavelength SNR catalogs in order to address the detection efficiency of SNR populations in different wavebands.

The outline of this paper is as follows: In \S2 we briefly describe
each galaxy and prior multiwavelength observations of their SNR
populations. In \S3 we describe the observations, the data reduction,
the analysis techniques used to detect SNR candidates and measurement of their X--ray properties. In \S4 we discuss the classification of the X--ray detected SNRs in this study and their correlation with already known X--ray SNRs. Our results are discussed in \S5 and finally in \S6 we present the conclusions of
this work.

\section{Sample selection}

The sample used for this study consists of six nearby galaxies: \objectname {NGC 2403}, \objectname {NGC 5204}, \objectname {NGC 4395}, \objectname {NGC 4449}, \objectname {NGC 3077} and \objectname {NGC 4214}.  These galaxies are selected from the Third Catalog of Bright Galaxies (RC3; \citealt{de Vaucouleurs95}) to be: a) late type (T$>4$; Hubble type), b) close ($\leq$ 5 Mpc) in order to
 minimize source confusion (at 5 Mpc, 1$\arcsec$$\simeq25$ pc), c) at
 low inclination ($\leq$ 60 degrees) in order to minimize internal
 extinction and projection effects and d) be above the Galactic plane ($\restriction$b$\restriction>20\degr$). From this pool of objects we selected galaxies which have {\it Chandra} archival data with exposure times long
 enough to achieve a uniform detection limit of ~10$^{36}$ erg
 s$^{-1}$. We opted to focus on {\it Chandra} data because of its superior spatial resolution which allows the detection of faint sources in crowded environments. The properties of our sample of galaxies are presented in
 Table 1.

\subsection{Previous surveys of SNR populations in our sample of galaxies}

Many of the galaxies in this sample have been extensively studied in several wavelengths. Next we summarize these results and we present any previous X--ray studies of their SNR populations.

{\it NGC 3077}: is a member of the gas-rich M81 group, where
interaction between M81 and M82 is believed to have triggered
starburst activity to both M82 and NGC 3077 \citep{Walter02}. In a
{\it Chandra} study of NGC 3077, \citet{Ott03} reported the detection of 3
X--ray sources (S1, S5 and S6) with X--ray characteristics indicative
of thermal SNRs. One of these sources (source S1) coincides with a
radio source detected by \citet{Rosa05}.
 
{\it NGC 4214}: is a nearby irregular starburst galaxy which features
extensive massive star formation throughout its disk. Therefore the existence of SNRs is expected. There is one radio source classified as a radio SNR (source - {\it $\varrho$} of \citealt{Vukotic05}) while the nature of sources {\it $\alpha$} and {\it $\beta$}) (also from the list of \citealt{Vukotic05}) is debated \citep{CW09}. In addition \citet{CW09} find 6 more radio candidate SNRs and 3 sources denoted as
SNR/H{\sc ii}. However, no X--ray emitting SNRs have been identified
so far in NGC 4214.

{\it NGC 4395}: is another irregular starburst galaxy that may host a
candidate radio SNR \citep{Sramek92, Vukotic05} but with no X-ray
counterpart. This source is outside the field of view of the observations studied in this work. Up to now, there are no other SNRs identified in NGC 4395 in any wavelength.

{\it NGC 4449}: This Magellanic-type irregular, starburst galaxy hosts
an extensively studied Cas--A like SNR (e.g. radio: \citet{Lacey07};
optical: \citet{Blair83}; ultraviolet \citet{Blair84}, and X--rays:
\citet{PF03}). In addition, \citet{CW09} detected 8 candidate SNRs
based on radio observations and H$\alpha$ images. In the X--ray band,
\citet{Summers03} report the presence of 2 SNRs
and 8 SNR/XRB or SNR/SSS (SSS: Super Soft Source) systems, based on {\it Chandra} data.

{\it NGC 5204}: Three SNRs were optically identified in this irregular
galaxy by \citet{MF97}. Observations for the detection of SNRs in any
other waveband have not yet been conducted.

{\it NGC 2403}: \citet{MFBL97} (hereafter MFBL) performed a relatively
deep, ground-based, optical search for SNRs in the Scd galaxy NGC
2403. They identified 35 SNRs, two of which were previously known
\citep{Dodorico80,Blair82}. \citet{TH94} classified two radio sources
(TH2, TH4) as SNRs, while \citet{Eck02} detected a
radio counterpart (denoted as source {\it $\mu$}) to the optically
identified SNR MFBL 7. \citet{SP03} and \citet{P07} searched for
positional coincidences between their sample of X-ray sources in NGC
2403 and the 35 SNRs of \citet{MF97}. They found one clear association
with MFBL31 as well as the candidate radio SNR TH2.

In Table 2 we summarize the numbers of optically identified SNRs, radio and already known X-ray SNRs for the galaxies in our sample, emerging from previous studies. Although these studies have presented candidate SNRs, their identification has been based mainly on their
multiwavelength associations. Here we present a data analysis focusing
on the identification of X--ray selected SNRs based on their thermal
X--ray emission.

\section{Data Reduction}

We have analyzed {\it Chandra} archival data for the six galaxies in our
sample. All exposures were obtained with the back-illuminated Advanced CCD
Imaging Spectrometer ACIS-S3 CCD chip (pixel size: 0$\farcs$49
$\times$ 0$\farcs$49; energy resolution $\thicksim$120 eV at 1keV;
\citealt{Garmire03}) at the focal plane of the High Resolution
Mirror Assembly \citep{van97}. The log of the observations is
presented in Table 3. We selected observations performed in FAINT Data Mode, full array and exposure times longer than 15 ksec. These parameters provide the largest field-of-view and ensure the detection of sources with X--ray luminosity as low as 10$^{36}$ erg s$^{-1}$ for the most distant galaxy. We found nine datasets for four galaxies of our sample (NGC 2403, NGC 3077, NGC 4214 and NGC 4449) that fulfill the above parameters. In the case of NGC 4395, three observations were available, fulfilling the FAINT Data Mode and exposure time criteria. Two out of the three were performed in 1/8 subarray mode while the third one was performed in custom subarray mode. We used only the third observation which covers a large area of the galaxy ($\thicksim$ 30\% of its $D_{25}$). As for NGC 5204, observations are performed in 1/8 subarray mode. Of these, 12 have short ($\backsim$ 5 ksc) exposures and only one has a longer 15 ksc exposure, which fulfills our exposure threshold. However, the 12 5ksc exposures are taken with different roll angles. Since each of them fulfill our 10$^{36}$ erg s$^{-1}$ limiting luminosity requirement and their combination covers $\backsim$ 80 $\%$ of the $D_{25}$ area of the galaxy, we opted to use them in our study. 

The data analysis was performed with the CIAO tool suite version 3.4 and
CALDB version 3.3.0, unless otherwise stated. Each dataset was analyzed following the {\it Chandra} Standard Data Processing threads (SDP)\footnote{See http://asc.harvard.edu/ciao/threads.}. 

Since we are interested in multi-wavelength comparison of the X--ray sources, application of astrometric correction is a critical element of this study. We searched for counterparts of point--like, bright X--ray sources ($>$ 100 counts) in the 2MASS All Sky Catalog of point sources (\citealt{Cutri03}). We found typically 3-5 counterparts for each galaxy. The comparison between the coordinates of the X--ray sources and the 2MASS counterparts yielded offsets $<$ 0.5 $\arcsec$ consistent with the average astrometric error of the {\it Chandra} Pointing\footnote{See http://cxc.harvard.edu/cal/ASPECT/celmon/}. We applied these offsets to the event files. Then, we reprocessed them to apply corrections for gain and Charge Transfer Inefficiency (CTI). We searched over source free regions for background flares by creating lightcurves with a time resolution of 200 sec. We applied a sigma--clipping algorithm\footnote{See http://cxc.harvard.edu/ciao3.4/ahelp/analyze\_ltcrv.html.} on the lightcurves in order to identify time periods of anomalous background levels ($\pm$ 3 $\sigma$ from the mean value). No significant flaring of the background was observed in any of the datasets.

\subsection{Imaging Analysis}

After the initial processing of the data, we created images in four
energy bands: soft (S: 0.3 - 1.0 keV), medium (M: 1.0 - 2.5 keV), hard
(H: 2.5 - 7.0 keV) and total (T: 0.3 - 7.0 keV). In the case of
multiple observations of an object, we aligned each event file of a given
observation to a chosen reference event file. We then reprojected and
merged the events of each band (S, M, H, T) for each dataset using the
CIAO script {\it mergeall}.

We set the upper energy limit of the medium band to 2.5 keV because energies up to that value include emission lines from hot
thermal plasma and are influenced by photoelectric absorption of the
column densities typically seen in galaxies. These emission lines are of great importance to investigate since they allow us to distinguish between low temperature thermal emitting gas and harder emission from a power-law or hotter thermal continuum. The goal of this study is the identification of X--ray emitting SNRs and our first classification criterion is based on the Soft X--ray color: Col1 = log(S/M), where S, M are the counts in the soft and medium bands. Therefore, we defined the energy range of the soft and medium bands in a way that provides the maximum separation of the S/M color distribution, for the optically thin, soft thermal plasma models characteristic of thermal SNRs (kT $\backsim$ 0.5 - 1.5 keV). We identified these bands by estimating the expected number of counts in different energy bands in a typical observation of our galaxies, assuming an apec (\citealt{Smith01}) thermal plasma model of different temperatures. The transition between the soft and medium band was in the 0.5 - 1.5 keV range since in this range the peak and strength of the FeL blend, which is characteristic of thermal plasma of this temperature, shows the strongest variation. 

 In Fig. 1 we show the S/M color for four different choices of the two bands: S$_{1}$: 0.3 - 0.5 keV and M$_{1}$: 0.5 - 2.5 keV, S$_{2}$: 0.3 - 0.7 keV and M$_{2}$: 0.7 - 2.5 keV, S$_{3}$: 0.3 - 1.0 keV and M$_{3}$: 1.0 - 2.5 keV, S$_{4}$: 0.3 - 1.5 keV and M$_{4}$: 1.5 - 2.5 keV. The different points show the S/M color which corresponds to different choices of temperature (0.25 - 2.0 keV from the bottom towards the top points of each set). As we can see from this figure, the third and fourth band selections (S$_{3}$: 0.3 - 1.0 keV and M$_{3}$: 1.0 - 2.5 keV, S$_{4}$: 0.3 - 1.5 keV and M$_{4}$: 1.5 - 2.5 keV) give the maximum discrimination of the S/M color for the different temperatures of a thermal model. Since these particular band selections do not show any major differences, we chose the third one (S$_{3}$: 0.3 - 1.0 keV and M$_{3}$: 1.0 - 2.5 keV; circles in Fig. 1) in order to ensure a larger number of counts in the medium band. Regarding the other energy bands, the combination of the medium and the hard band (2.5 - 7.0 keV) gives a good representation of the continuum emission and temperature while the total band (0.3 - 7.0 keV) is very useful for measuring the total flux of a source especially in the case of a small number of counts (e.g. \citealt{Zezas06}).

In order to directly compare the intensity and spectra of sources
detected by different instruments or observations, we accounted for
variations of the ACIS sensitivity by creating exposure maps for each
data set in the four bands (S, M, H and T). The CIAO 3.4 suite and CALDB 3.3.0
allow us to include in the exposure maps the time-dependent spatial
variations of the ACIS sensitivity due to the contaminant on the
detector window. Each energy's band exposure map is the sum of subbands (monochromatic maps), weighted only for the differences in the bandwidth. This is equivalent to a flat energy spectrum ($\Gamma$=0, N$_{\rm H}$=0) or estimating the integral of the effective area over each broad band \citep[see][]{Zezas06}. In the case of single observations we used the CIAO tool {\it mkexpmap}.

We note that because {\it mergeall} for a multiple--chip, multiple observation exposure map was not
able to extract merged exposure maps or use correctly the asol files
of the multiple exposures, we used the same script for extracting
exposure maps for each band and each observation and then combined
them with the CIAO tool {\it dmimgcalc}.

\subsection{Source Detection}

We searched each dataset for sources in the four energy bands, using
the {\it wavdetect} tool of CIAO\footnote{See
http://asc.harvard.edu/ciao/threads/wavdetect.}. The significance
threshold was set to one false detection over the searched area while the scales parameter was set to 2.0, 4.0, 8.0 and 16.0 pixels. We used the relevant exposure maps in order to avoid the detection of spurious sources close to the CCD edges. The results of the {\it wavdetect} run for each band and for each exposure were
cross-correlated and combined to create one source list for each
galaxy, covering the $D_{25}$\footnote{The $D_{25}$ ellipse is defined
as the optical isophote at the B--band surface brightness of 25 mag
arcsec$^{-2}$ (e.g. RC3, \citealt{de Vaucouleurs95}).} ellipse of the galaxy. In the case of
multiple observations, the detection was performed on the co-added
images, in order to achieve the maximum sensitivity.

We detected a total of 244 discrete X--ray sources in our sample of
six galaxies (22 in NGC 3077, 16 in NGC 4395, 26 in NGC 4449, 44 in NGC 4214, 125 in NGC 2403 and 11 in NGC 5204) down to a limiting flux of 10$^{-15}$ erg cm$^{-2}$s$^{-1}$ in the 0.3 - 10.0 keV band.

We calculated the expected number of background sources for each galaxy in our sample based on the {\it Chandra} Multiwavelength Project (ChaMP) X--ray point source catalog and the cumulative luminosity distribution of \citet{Kim07}. Given that the investigated sources of this study are predominantly soft sources, we used the 0.5 - 2.0 keV logN - logS in order to estimate the number of background sources down to our limiting flux of  10$^{-15}$ erg s$^{-1}$ cm$^{-2}$ within the area of each galaxy covered by the {\it Chandra} observations. The number of expected background sources is given in Table 4.

\subsection{Photometry}

In order to perform photometry of the detected X--ray sources, we defined
source apertures on the total band image, while ensuring that i) they do not encompass other
neighbouring sources or significant diffuse emission and ii) they cover at
least 90\% of the encircled energy of the source's Point Spread Function (PSF) at a given off--axis angle and a typical energy of 1.4 keV (appropriate for the soft sources we investigate in this work). The typical radius of these apertures is $\backsim$ 1.0$\arcsec$ - 1.5$\arcsec$.  The background for each source was measured locally from apertures defined to cover a large area around them, while ensuring that the diffuse emission does not vary significantly
over the background area. We measured the number of raw counts for each source
and its corresponding background in each of the S, M, H and T bands, using the {\it
dmextract} tool of CIAO\footnote{See
http://asc.harvard.edu/ciao/threads/dmextract.}. To account for
exposure variations across the ACIS-S3 CCD area, we calculated corrections of the effective area for each source
by normalising the exposure maps of each band with respect to a 
reference point close to the center of each galaxy such as to minimize variations of the sensitivity over the studied area (see \citealt{Zezas06}). In the case of
multiple exposures, the reference point was chosen on one of the
merged exposure maps. This way we also correct for sensitivity variations between different observations. In Tables 5-10 we present the observed (raw)
source and background counts in the S, M, H and T bands for sources
that appear to be potential SNRs (see \S 3.4.1) and sources with very strong soft components based on their X--ray colors. We also present the background/source area ratio (which is common in all
energy bands since the same source and background apertures were used) as well as the background/source effective area ratio in the soft energy band. 

\subsection{Spectral analysis}

\subsubsection{X--ray colors}

Since our goal is to identify candidate sources with soft thermal
spectra (kT $\le$ 2 keV, typical of thermal SNRs; Schlegel 1994),
initially we use the X--ray photometry derived in the previous
section. Instead of hardness ratios which have less symmetric
posterior probabilities \citep{Park06}, we calculated X-ray colors
defined as C1=log(S/M), C2=log(M/H) and C3=log(S/H), where S, M, H are
the net counts in the Soft (0.3 - 1.0 keV), Medium (1.0 - 2.5 keV) and
Hard (2.5 - 7.0 keV) band, respectively. The X--ray colors can be used
to obtain information on the spectral properties of the X-ray emission
even in the case of small number of counts when spectral fitting is
not possible.

In the small number of counts regime where most of our sources belong,
the Poisson distribution becomes distinctly asymmetric. In this case
it is more appropriate to use a method based on the Bayesian estimate
of the "real" source intensity which takes into account the Poisson
nature of the probability distributions for the source 
counts as well as the effective exposure at the position of the source
\citep{van01, Park06}. Furthermore, in the case of non-detections in
one or more bands this method can also provide upper limits on the X--ray colors. Since the
numerical integration of the Bayesian method is more accurate but
computationally more intensive, we used it for sources with fewer than
70 counts in any of the soft, medium or hard bands. The integration was performed using the Gaussian quadrature
algorithm with 2500 bins. For sources with more than 70
counts, we used the Gibbs algorithm with 10$^{5}$ draws, 15000 of which were rejected as burn--in
draws. In both cases the confidence level was set at 68.0$\%$.

 In Figs. 2--7 we plot the C1 against C2 X-ray colors for the detected
 X-ray sources. On the same plots we added grids for power-law and
 thermal plasma models for different values of temperature (kT), absorbing H{\sc
 i} column density (N$_{\rm H}$) and photon index $\Gamma$. Sources that
 appear to have temperatures below 2 keV and mainly lie on the thermal
 grid (locus of SNRs) as well as those which are consistent with the
 thermal grid within their error--bars, are potential SNRs. In the case
 of plerion SNRs, we expect them to lie on the power--law grid as their
 X--ray emission is non--thermal with photon indices of 1.7--2.0 (e.g \citealt{Asaoka90}). Since their spectra are very
 similar to those of X--ray binaries, they cannot be selected solely
 on the basis of their X--ray properties. In Tables 11-16 we present the
 calculated X--ray colors (C1, C2 and C3) for potential SNRs in all
 galaxies of our sample as well as for the soft sources that lie on
 the right corner of the plots which indicate an extra soft component
 due to diffuse emission. For comparison, we show which sources have
 extracted spectra (see \S3.4.2) and the suggested classification based on their spectroscopy (see $\S$4).

\subsubsection{Spectral fitting}

We tested the X--ray color tentative classification of thermal SNRs by performing
spectral analysis for sources with adequate number of counts. We also
examined the soft sources on the bottom right corner of the color--color
plots. We extracted PI spectra, auxiliary response files and redistribution matrix files with the {\it specextract CIAO} script which takes
into account spatial variations of the effective area by creating the
weighted redistribution matrix files and auxiliary response files
(wrmf, warf) for each source. Spectral fit was performed with XSPEC
version 11. The upgraded 3.4.2 version of CALDB was used.

For sources with more than 50 counts, $\chi ^{2}$ statistics were
used. The spectral channels were binned for the spectral fitting
analysis in order to contain at least up to 25 counts per bin before
background subtraction for sources with high background (source counts with $>$ 5 $\%$ contamination from background counts) and 15 counts for
sources with low background ($<$ 5 $\%$ contamination from background counts). This ensures that in either case we have
between at least 10 counts after background subtraction in each bin. The corresponding background was
subtracted from the source spectrum during the fitting process. For
sources with few counts ($\la 50$) the Cash maximum likelihood
statistic was used \citep{Cash79}, which is more appropriate than
$\chi ^{2}$ statistics in the case of small number of counts. The
spectra were not binned in order to preserve the maximum amount of information for
statistical analysis.

Each spectrum was initially fitted separately with two different
models: power--law (PL) and thermal plasma (apec;
\citealt{Smith01}). We opted to use the apec model since it is the most up to date optically thin equilibrium thermal plasma model including significant improvements in the number of spectral lines and their oscillator strengths. In a few cases where we had indication for abnormally strong emission lines, indicative for a non-equilibrium plasma, we also used the nei model (e.g. source LZB15 in NGC 3077). Also sources well-fitted with an apec model but presented unrealistically low temperatures for a thermal plasma ($<$ 0.1 keV), were re-fitted with a blackbody (bbody) model.

 The models were coupled with a photoelectric
absorption model (phabs) due to the Galactic Interstellar Medium
as well as to the material within each galaxy, with the restriction
that the derived column densities should exceed the weighted average
Galactic line-of-sight values obtained from the LAB
(Leiden/Argentine/Bonn) Survey \citep{Kalberla05} of Galactic H{\sc i} \footnote{See http://heasarc.nasa.gov/cgi-bin/Tools/w3nh/w3nh.pl}. We also assumed
solar metallicities for all galaxies since the quality of the data did
not allow us to constrain the abundance.

In some cases (e.g. LZB4 and LZB5 in NGC 4449; LZB24 in NGC 4214; LZB64, LZB99 and LZB101 in NGC 2403; LZB11 in NGC5204) the single component models gave either unacceptable fits ($\chi ^{2}$ $>$ 2), strong line-like residuals or unrealistic best fit parameters. In this case we fitted the data with a two-component (PL + apec) model with the same absorption. As discussed in \citet{Protassov02}, the use of the F-test in order to assess the improvement of the fit in this case is not statistically proper.

We selected the best model for each source based on the
quality of the fit (e.g. good statistical fit mainly determined
by $\chi_{\nu}$$^{2}$$\thicksim$1 for chi-square statistics, and
goodness-of-fit $\thicksim$50$\%$ for Cash statistics) and the
plausibility of the parameter values (thermal component temperature $\le$ 3 keV, power law photon index $\le$ 4). In the case of multiple observations, we fitted the individual spectra simultaneously with the same model. All model parameters apart from the normalization were determined by the first dataset.

The best fits obtained for potential SNRs (sources in the locus of
SNRs as well as those in the bottom right corner of the color-color plots) in
each galaxy are summarized in Tables 17-22. Column 1 presents the
source ID as well as the observation ID in the case of multiple
exposures. For the latter, we used only observations at which the
corresponding source is detected with adequate number of counts and does not lie near
the borders of the CCD ($\thicksim$100 pixels away). Column 2 shows 
the fitted model. Column 3 gives the H{\sc i} column
density obtained from fitting the phabs (photoelectric absorption)
component and Column 4 gives the best fit photon index for the power--law model or
the temperature for the apec or black body models. Column 5 presents
the model normalization. In the case of multiple exposures, we give the multiplicative factor for the intensity of the different spectra that were fitted simultaneously, with respect to the first observation of each source. All uncertainties at the 90$\%$ confidence level are
calculated with the {\it error} command of XSPEC. Columns 6 and 7
present the absorbed and unabsorbed flux in the 0.3-10.0
keV energy band. Although spectroscopy is derived from the 0.3-7.0 keV
energy band, we chose to calculate the fluxes in the 0.3-10.0 keV
energy band for consistency with previous X--ray publications. Column
8 is the source classification based on the X--ray spectral fits: the presence
of only low temperature thermal component(s) in a fit strongly indicates that the source
is an SNR while sources having hard components (power--law) to their
spectra are most likely XRBs. Sources well--fitted with a low--temperature ($\thicksim$100 eV) black body model are denoted as Super Soft Sources (SSS).

We mention that there are cases of SNRs (e.g. LZB15 in NGC 3077) where
the electron temperature of the best--fit thermal model is relatively
high ($>2$ keV). This may imply a non-equilibrium state of the
collisionally ionized plasma of the source (these sources are also
well-fitted with a nei model). However, this is not an unusual
phenomenon. \citet{Kong04} propose that non-equilibrium state of the
collisionally ionized plasma could either come from the
shock--heated swept--up circumstellar medium or it is due to
inhomogeneity of the ISM. 

What is more, there are sources with a small
number of counts that are fitted with high values of column density.
There is a well known positive correlation between the
normalization of a model and the absorbing column density. Therefore,
in the case of low number of counts spectra (such as those we
consider here) it is possible that we can obtain high best-fit
H{\sc i} column densities and hence high inferred unabsorbed source
luminosities (resulting from the high normalization of the low-energy
source spectral components).

\section{SNR classification and correlations with already detected X--ray SNRs}

On the basis of these results, we divide our X-ray detected SNRs into
three types: a) SNRs, b) probable SNRs, and c) candidate SNRs. As SNRs
we consider point--like non--variable sources which have spectra consistent with a
single or two--component low temperature thermal X--ray spectrum (kT$<$3 keV). We consider 
as probable SNRs sources which: i) fulfill the above criteria but have a small number of counts ($<$50) and/or
 large errors on their spectral parameters and/or ii) vary by $< 15\%$ in flux between different
observations. Candidate SNRs are sources for which it was not possible
to extract any X--ray spectra but fulfill the hardness ratio criteria
(i.e. they lie, within their error bars, on the low--temperature part of the thermal grid of the color--color plot). Sources for which 
we cannot distinguish between a thermal and a non--thermal model are denoted as unclassified. Nonetheless, their spectral parameters are also
presented in the relevant tables.

We find good agreement between the spectral parameters derived from the
X--ray colors and the analysis of the X--ray spectra, giving us
confidence in the use of  hardness ratio diagrams as a diagnostic tool for the
 initial identification of thermal SNRs. We note that all sources
considered in this study are pointlike (physical scales 5 and 11 pc for
the closest and more distant galaxies respectively) limiting the possibility that
they are local enhancements of the general diffuse X--ray components.

A total of 37 X--ray selected SNRs (8 SNRs, 24 probable SNRs and 5 candidate SNRs) are detected in this study, 30 of which are new identifications. One third of these new sources have been also identified as SNRs based on other multi--wavelength observations, giving us confidence that the X--ray selection scheme is robust. We are in the process of analysing additional deeper multi--wavelength data in order to extend the classification to other objects (Leonidaki et al. in preparation). Three of the galaxies in our sample (NGC 4214, NGC 5204 and NGC 4395) containing 71 X--ray detected sources are studied for the first time in this context, and they exhibit 12 new X--ray SNRs. There is one additional source in NGC 3077 which appears to be in the data after visual inspection of the raw and smoothed images. This source seems to be extended and in a location with significant diffuse emission, and is not detected by wavedetect. The same source was also included in the source list of \citet{Ott03} only after visual inspection of the data (source 5, classified as SNR). Since it could be a local enhancement of diffuse emission and for that reason has not been detected by a robust blind search algorithm like wavedetect, we opted not to include it in our source list. No other X-ray known SNRs have not been detected by this study.

In the case of two sources in NGC 4449 our results suggest slightly different classifications than those published in previous studies (e.g. \citealp{Summers03}). We classify source LZB4 as an XRB based on its relatively hard spectrum ($\Gamma$=2.3). Source LZB26 can be fitted equally well with an absorbed apec (kT$\thicksim$1.02 keV) or an absorbed black body (kT$\thicksim$0.2 keV). Therefore we consider it as a probable SNR.

The requirement that the sources we consider as SNRs are non--variable minimizes the possibility that they are super--soft or quasi--soft X--ray sources. Furthermore, super--soft sources typically have much softer spectra (kT $\leq$ 0.1 keV), while quasi--soft sources have soft spectra with a power--law component (e.g. \citealt{Greiner96, Di Stefano04}). Our SNRs instead have typically ``clean'' thermal spectra. Additionally, one third of our X--ray SNR identification are confirmed based on known SNRs in other wavelengths, giving us confidence on these selection criteria.
We cross-correlated all X-ray selected SNRs against the 2MASS catalog and checked their optical counterparts on SDSS images. 
We found only one source (LZB22 in NGC4449) that coincides with a foreground star.

Furthermore, the soft thermal spectra (kT $\leq$ 2 keV) of the detected SNRs minimize the possibility that they are background sources such as AGNs or QSOs. The latter exhibit harder X--ray spectra with photon indices $\Gamma$ $\leq$ 2 - 2.5, added as such they would have been excluded in our selection process. 

We also cross-correlated all X-ray selected SNRs against the 2MASS catalog and checked their optical counterparts on SDSS images. Only one source (LZB22 in NGC4449) coincides with a foreground star. This source is not included in the SNR sample of the relevant galaxy.

\section{Discussion}

\subsection{Multiwavelength associations}

Comparison of the emission of SNRs in different wavebands can provide information about the evolutionary stage of the sources and/or can illustrate selection effects. For that reason we searched for coincidences between the X--ray identified SNRs (see $\S$4) with SNRs in optical and radio wavebands (see $\S$3). In Fig. 8 we present the overlap between X--ray, optical and radio selected SNRs, in the form of Venn diagrams for all galaxies in our sample, except for NGC 5204 where no X--ray SNRs were identified. The X--ray sources we consider in this comparison are denoted as SNRs or probable SNRs. All multi-wavelength comparisons were performed for the same area for each galaxy. For that reason we excluded the radio SNR in NGC 4395 \citep{Vukotic05} as it is outside the {\it Chandra} field of view. In addition, we excluded the radio candidate SNRs $\alpha$ and $\beta$ in NGC 4214 \citep{Vukotic05} the nature of which is debated \citep{CW09}. In this comparison, we consider only radio candidate SNRs, excluding SNR/H{\sc ii} composite objects from \citet{CW09} which present spectral index consistent either with an H{\sc ii} region or SNR. 

From the 36 optically identified SNRs (mainly on the basis of narrow--band photometry) 8 possess X--ray counterparts (corresponding to a detection rate of 22$\%$), while 7 out of the 19 radio--candidate SNRs have X--ray counterparts (detection rate of 37$\%$). Little overlap appears to be between optical and radio SNRs ($\thicksim$6$\%$). In the case of NGC 2403 (Fig. 9) we find a larger number of X--ray SNRs than reported in the study of \citet{P07}. This is due to: (a) the much longer exposures used in the present study, and (b) the different selection criteria (\citealt{P07} focused on optically/radio selected samples of X--ray emitting SNRs).

The detection rate of SNRs in different wavebands strongly depends on the properties of the surrounding medium of the source. For example, \citet{P07} point out that optical searches are more likely to detect SNRs located in regions of low diffuse emission, while radio and X--ray searches are more likely to detect SNRs in regions of high optical confusion. In this study, the sample of radio SNRs is limited by the lack of deep radio surveys for SNRs for half of our galaxies. This could contribute to the difference in the detection rates between optical/X--ray SNRs and optical/radio, X--ray/radio SNRs.

\subsubsection{Supernova Remnants or X--ray Binaries?}

Three X-ray sources (LZB93, LZB99, LZB104) in NGC 2403, although spectroscopically identified as XRBs on the basis of their hard X--ray emission (and with X--ray luminosities consistent with those of XRBs, see Table 23), are associated with optically known SNRs in the catalog of \citet{MFBL97}. This is also the case for the spectroscopically identified XRB (LZB26) in NGC 4214 which coincides with a radio SNR/H{\sc ii} source detected by \citet{CW09} (see Table 23). One possible interpretation is that of an X--ray binary coincident with a supernova remnant, possibly associated with the supernova that produced the compact object in the binary. In this case the SNR is responsible for the observed optical and radio emission while the binary system produces the X-ray emission. The X--ray luminosity of active XRBs ($\ge$10$^{37}$ erg s$^{-1}$) is higher than that of SNRs (typically 10$^{35}$ -- 10$^{37}$ erg s$^{-1}$) and therefore they can overshadow the latter. The exemplar of this type of objects is the SS 433/W50 SNR/XRB system (e.g. \citealp{Safi99}), while a few other candidates have been identified in other galaxies on the basis of hard and/or variable X--ray sources associated with optically or radio identified SNRs \citep{P07}.

We searched for additional sources of this class witnessed by composite thermal -- non-thermal spectra (point-like sources requiring power--law and apec spectral components) where the thermal component dominates, indicating that they could be SNR/XRB systems. Four X-ray detected sources in our sample of galaxies (LZB4 in NGC 4449, LZB24 in NGC 4214 and LZB64, LZB101 in NGC2403) are fitted with  composite thermal -- non-thermal spectra. The thermal component contributes $<$ 30$\%$ to the total X-ray emission of these systems therefore their classification as XRBs is more robust.

\subsection{N$_{\rm H}$ - $L_{X}$, kT - $L_{X}$ }

We examine the H{\sc i} column density (N$_{\rm H}$) of the X--ray selected SNRs as a proxy of the density of their local ISM. SNRs in dense star--forming regions are usually associated with significant amounts of cold gas, which could result in excess absorption towards their line-of-sight. Simulations of SNRs embedded in dense environments show that they tend to have higher luminosities (e.g \citealt{Chevalier01}). Therefore, one would expect a correlation between their luminosity and the density of their environment.
In Fig.10 we plot the column density against the absorption--corrected luminosity of the spectroscopically identified SNRs in this study, based on their best fit parameters (Tables 17-22). The Galactic column density value has been subtracted from the measured N$_{\rm H}$ value. The error bars correspond to the 90$\%$ confidence level for one interesting parameter. In this plot we do not include sources with N$_{\rm H}$ value peged on the Galactic value. Non--existence of down--side N$_{\rm H}$ or left--side luminosity error bars indicates upper bounds at the 90$\%$ confidence level. In the same plots we also show for comparison a sample of Magellanic Cloud (MC) SNRs from the {\it Chandra} Supernova Remnants Catalogue\footnote{See http://hea-www.cfa.harvard.edu/{\it Chandra}SNR/}. From that plot we see that our sample of extragalactic SNRs have systematically higher X--ray luminosities than the MC--SNRs and that there is not a trend between H{\sc i} column density and luminosity. This could indicate that the contribution of the local environment to the overall H{\sc i} column density is very smaller, or that the local densities are not high enough to significantly affect the X--ray luminosity.

In Fig. 11 we investigate the correlation between SNR temperature and absorption--corrected luminosity. The majority of our SNRs have temperatures in the 0.1--1.0 keV range, typical for thermal SNRs (e.g. \citealt{S94}) and luminosities in the 5$\times$10$^{36}$ to 5$\times$10$^{39}$ erg s$^{-1}$ range. We do not see any significant correlation between luminosity and temperature. However, we do see a population of SNRs with high temperatures or high luminosities. As discussed in \S3.4.2 the objects with high temperatures indicate sources with non--equilibrium spectra while the high absorption--corrected luminosities of few objects are probably artifact of their large (and often poorly constrained) column density. In fact often the luminosities of these objects are consistent with the main population of SNRs within their errobars. The dashed line indicates the expected relation between the unabsorbed X--ray luminosity and temperature for a thermal source at a distance of 5 Mpc, based on an apec model with fixed emission measure (EM). From this model we see a weak dependence of the X--ray luminosity to temperature if it is below 1 keV. We note that in the case of double thermal models (apec+apec), the higher value temperature was used.

\subsection{SNRs and Star Formation Rate (SFR)}

Since core--collapse SNRs are the endpoints of the evolution of the most massive stars, they are good indicators of the current SFR. Although there are several calibrations of the radio SNR rate - SFR (e.g. \citealt{CY90}), the investigation of the X--ray SNR rate with SFR are hampered due to the lack of large and secure samples of X--ray selected SNRs. Here we attempt to derive such a calibration using the uniform samples of X--ray SNRs selected in \S4.

All galaxies in our sample have accurate measurements of their integrated FIR luminosity (\citealt{Ho97}, Table 25), so we opted to use this as a SFR proxy. The FIR luminosity is based on integrated flux measurements with IRAS in the 60 and 100 $\mu$m bands which were used to calculate the 42--122 $\mu$m broad--band FIR luminosity using the calibration of \citet{Rice88, Helou88} and the distances in Table 1. We are aware that the FIR luminosity tends to overestimate the instantaneous SFR since it includes contribution from late type stars, however for the purpose of the comparison of the SNR rate with the SFR in this set of galaxies it is better suited than the H$\alpha$ luminosity which is heavily affected by extinction.

\subsubsection{X-ray properties of SNRs and Star Formation}

In order to investigate the X--ray properties of SNRs in different star--forming environments, we calculate the average integrated 0.3 - 10.0 keV unabsorbed X--ray luminosity of SNRs detected down to a luminosity of 5$\times10^{36}$ erg s$^{-1}$ for each galaxy. This limiting luminosity is based on a preliminary analysis of the luminosity distribution of the SNRs in our sample (Leonidaki et al. in preparation) which shows that the sample of SNRs is complete down to this limit. We only consider sources classified as SNRs or probable SNRs by this study. We also include in our sample the X--ray detected SNRs (above our detection limit) in three more galaxies from the work of \citet{P07}. The 0.2-10.0 keV average unabsorbed X--ray luminosity of the SNRs in Pannuti et al. were converted to the 0.3-10.0 keV energy range assuming a thermal bremsstrahlung model with a temperature of kT = 0.5 keV. Although the 0.1 keV difference in the energy bands has only a small effect ($\thicksim 6.5 \%$) on the unabsorbed X--ray luminosity for a typical SNR spectrum, we performed the conversion for consistency. 

In Fig. 12 we plot the average unabsorbed X-ray luminosities of SNRs in spiral galaxies (squares) and irregular galaxies (triangles) from our sample against the integrated 42--122 $\mu$m FIR luminosity of each galaxy which is a good star-formation rate indicator. In the same plot we include the SNRs with luminosity greater than 5$\times$10$^{36}$ erg s$^{-1}$ SNRs (circles) from the LMC and SMC ({\it Chandra} Supernova Remnants Catalogue).

As expected, we do not see any correlation between the average SNR X--ray luminosity and the total FIR luminosity of their host galaxy. However, we do see a systematic trend for more luminous SNRs to be associated with irregular galaxies. This indicates a difference of the SNR population characteristics between the two samples. This could be due to the typically lower metallicity of irregular galaxies than in typical spiral galaxies (e.g. \citealt{Pagel81}; \citealt{Garnett02}). Low abundances result in weaker stellar winds (e.g. \citealt{Lamers99}) which in turn  produce higher mass supernova progenitors. More massive progenitors are expected to produce more massive ejecta and stronger shocks which would lead to higher SNR X--ray luminosities.

Other possible interpretations include the non--uniform ISM which is often the case in irregular galaxies, or possible Initial Mass Function (IMF) differences between spiral and irregular galaxies. In the first case local enhancements of the ISM (especially at the star forming regions) could result to more luminous SNRs, while in the second case, flatter or top heavy IMFs which in some instances have been proposed for irregular galaxies, would result to larger number of SNe with more massive progenitors.

\subsubsection{Number of SNRs and SFR}

Since SNRs are the short--lived end--points of young stars we would expect a linear relation between the number of X-ray selected SNRs and SFR (e.g. \citealp{CY90}). To verify this connection, we plot the number of SNRs above the completeness limit of our sample (5$\times$10$^{36}$) against the integrated FIR luminosity of each galaxy (Fig. 13). For comparison, in the same plot we include the X--ray selected sample of SNRs in the MCs from the XMM-Newton study of \citet{Ghavamian05}. Although in the case of SMC this study covers a fraction of the area of the galaxy, a comparison with the ASCA--selected SNR sample of \citet{Yokogawa00} which covers almost the whole galaxy show that the sample of Ghavamian et al. is complete, and the MC SNR census extends down to a luminosity of 10$^{35}$ erg s$^{-1}$, much lower than our completeness threshold. 

In order to compare the Magelanic Cloud SNR populations with our sample, we use the luminosity distributions of the MCs SNRs from \citet{Ghavamian05} and rescale their observed numbers to the numbers of SNRs down to our limiting luminosity of 5$\times$10$^{36}$ erg s$^{-1}$, assuming a cumulative slope of $- 0.5$ (see Table 24), which is a good representation of the MC SNR populations. We did not use the SNRs in the three galaxies of Pannuti et al. since the different selection criteria of the two studies (optically selected X--ray emitting SNRs in the study of Pannuti et al. versus X--ray selected SNRs in this study) do not allow a direct comparison of the two populations.  

We find a linear relation between the number of X--ray selected SNRs and the FIR luminosity (Fig. 13), but the small number of objects does not allow us to quantify their scaling relation. However, a linear correlation coefficient of 0.72 shows that this is a significant correlation. Even if we remove NGC 2403, which seems to drive the correlation, or NGC 4449 which has the greatest distance uncertainties and hence FIR luminosity uncertainties, we measure a correlation coefficient of 0.53 and 0.90 respectively. 

The non-thermal radio emission is a more direct indicator of the supernova rate and hence high--mass star formation (e.g. \citealt{CY90}). Therefore we also investigate the correlation between the 1.4 GHz radio emission of the galaxies in our sample with the detected number of X--ray SNRs (Fig. 14). We use integrated radio fluxes from \citet{Condon87} and we find a correlation coefficient of 0.45. For the same reasons described in the previous paragraph, if we remove NGC 2403 or NGC 4449 the correlation coefficient is 0.22 and 0.85 respectively. The weaker correlation between the number of SNRs and the radio 1.4 GHz luminosity could be due to a significant contribution of thermal radio emission to the 1.4 GHz luminosity.

\subsection{Luminosity distribution of SNRs}

In order to examine our results in the context of SNR populations detected in other galaxies, we compare the luminosity distributions of X--ray SNRs in different types of galaxies with the number of X--ray detected SNRs in the studied sample. Therefore we test if the numbers of SNRs in the irregular galaxies in our sample are consistent with those expected, by simply rescaling the SNR X--ray luminosity Functions (XLFs) of the MCs \citep{Ghavamian05}.

We can estimate the expected number of SNRs in the galaxies of our sample based on a Magelanic Cloud - like SNR luminosity function and linear scaling of their number with star-formation rate. We find that the number of the observed SNRs in most galaxies of our sample is consistent with those expected by rescaling the MC XLF (Table 25), which has a cumulative slope of -0.5 \citep{Ghavamian05}. However, we see a large discrepancy between the observed and expected number of SNRs in NGC 4449 and NGC 2403. In the case of NGC 4449, this discrepancy could be due to the significant distance uncertainty to this galaxy which ranges between 2.9 \citep{Karachentsev98} and 5.0 Mpc \citep{Aaronson83}. The lowest distance value of 2.9 Mpc has been used in previous X--ray SNR surveys (e.g. \citealt{Summers03}) and the corresponding expected number of SNRs (6.75) is in fair agreement with the observed number (3). A more reliable measurement of its distance based on the TRGB (Tip of the Red Giant Branch) method (4.2 Mpc; \citealt{Annibali08}) results in 13.8 SNRs (see Table 25). Given the large uncertainties in its distance we exclude NGC 4449 from this comparison.

In the case of NGC 2403, the discrepancy of the observed population of X--ray selected SNRs with these expected based on the MC XLF and a linear scaling with SFR, may indicate a difference between their populations. Such a difference might be expected given that NGC 2403 is a grand design spiral galaxy, while the MCs are irregular galaxies. Therefore we use, instead of the MCs, the XLF of SNRs of M33 as our benchmark \citep{HaP01}. Since this XLF is not based on X--ray selected SNRs but instead optically/radio selected SNRs, for consistency we compare them with the SNR census of Pannuti et al. which is based on similar selection criteria. This also allows us to extend the comparison to a larger number of spiral galaxies. We follow the same approach as for the irregular galaxies with the only exception that we rescaled the number of M33 SNRs to the limiting luminosity of \citet{P07} (10$^{37}$ erg s$^{-1}$). We find a good agreement between the observed numbers of optically/radio selected SNRs in these spiral galaxies and the ones expected based on an M33--like XLF (Table 26).

Overall, our comparisons of the luminosity distributions of X--ray selected SNRs in different types of galaxies with the number of X--ray detected SNRs in the studied sample is suggestive of different luminosity distributions between the SNRs in spiral and irregular galaxies. This will be further examined by the comparison of the XLFs between SNRs in the spiral and irregular galaxies of our sample (Leonidaki et al. in preparation).

\section{Conclusions--Summary}

1. In this paper we have presented a systematic study of X--ray emitting SNRs in a sample of six nearby galaxies. The SNRs are selected on the basis of their soft X--ray spectra (kT $<$ 2 keV) or colors. We find a total of 37 X--ray SNRs, 30 of which are new identifications. Many of these SNRs are also detected in other wavebands which indicates that the X--ray colors are a good diagnostic for the primary identification of thermal SNRs. From the analysis of the sample we find: 22$\%$ of the optically identified SNRs and 37$\%$ of the radio candidate SNRs have X--ray counterparts . There is little overlap (5$\%$) between optical and radio classifications, which could primarily be due to the poor sensitivity of the existing radio surveys. 

2. Four sources identified as SNRs in optical or radio observations exhibit X--ray properties more consistent with XRBs. The latter and SNRs can be naturally associated since they are both related to short-lived high mass stellar objects, therefore we propose that these sources are X--ray binaries coincident with a supernova remnant.

3. We do not find any trend between the X--ray luminosity of SNRs and their H{\sc i} column density or temperature.

4. We find that X--ray SNRs in irregular galaxies appear to be more luminous than those in spiral galaxies. We attribute this effect either to the lower metallicity of irregular galaxies (which result in more massive progenitors) or their clumpy ISM.

5. We find evidence for a linear relation between the number of luminous X--ray SNRs ($L_{x}$ $>$5$\times$10$^{36}$ erg s$^{-1}$) and star--formation rate in our sample of galaxies.

6. There is a suggestion for different X--ray luminosity functions between the SNR populations of irregular and spiral galaxies, based on comparison of the observed numbers of SNRs and those expected by rescaling the luminosity functions of SNRs in the MCs .

\acknowledgments

The authors would like to thank John Raymond for fruitful discussions. This work was partly supported by NASA grant GO6-7086X and NASA LTSA grant G5-13056. IL and PB would like to thank the Harvard--Smithsonian Center for Astrophysics for its hospitality during their visits there. IL acknowledges funding by the European Union and the Greek Ministry of Development in the framework of the programme `Promotion of Excellence in Research Institutes (2nd Part).

\begin{deluxetable}{ccccccccccc}
\tablecolumns {11}
\rotate
\tabletypesize{\scriptsize}
\tablecaption{Properties of our sample of galaxies}
\tablewidth{0pt}
\tablehead{
\colhead{Galaxy} & \colhead{RA} & \colhead{DEC} & \colhead{Distance} & \colhead{Major and Minor axis} &\colhead{Inclination}\tablenotemark{a} &\colhead{Galactic N$_{\rm H}$}\tablenotemark{b} &\colhead{Galactic latitude} &\colhead{Position Angle}\tablenotemark{c} & \colhead{Type} & \colhead{Physical scale}\tablenotemark{d} \\
\colhead{} & \colhead{(J2000)} &\colhead{(J2000)} &\colhead{(Mpc)} &\colhead{(arcmin)} &\colhead{(degrees)} &\colhead{($\times 10 ^{20}$ cm $^{-2}$)} &\colhead{(degrees)} &\colhead{(degrees)}  &\colhead{} &\colhead{(pc)}} 
\startdata
NGC 2403 &07:36:51.4 &65:36:09 &3.2\tablenotemark{e}  &21.9{$\times$}12.3 &62       &4.36  &29    &127      &SAB(s)cd      & 7.7\\
NGC 5204 &13:29:36.5 &58:25:07 &4.8\tablenotemark{e}  &5.0{$\times$}3.0   &53       &1.75  &58    &5        &SA(s)m        & 11.5\\
NGC 4395 &12:25:48.9 &33:32:48 &2.6\tablenotemark{f}  &13.2{$\times$}11.0 &38       &1.85  &82    &147      &SA(s)m        & 6.2\\
NGC 4449 &12:28:11.9 &44:05:40 &4.2\tablenotemark{g}  &6.2{$\times$}4.4   &56       &1.61  &72    &45       &IBm           & 10.1 \\
NGC 3077 &10:03:19.1 &68:44:02 &3.6\tablenotemark{h}  &5.4{$\times$}4.5   &\nodata  &5.05  &42    &45       &I0 pec        & 8.6 \\
NGC 4214 &12:15:39.2 &36:19:37 &4.7\tablenotemark{i}  &8.5{$\times$}6.6   &37       &1.99  &78    &\nodata  &IAB(s)m       & 11.3\\
\enddata
\tablecomments{Units of right ascension are hours, minutes and seconds and units of declination are degrees, arcminutes and arcseconds. We note that the SA(s)m galaxies are Magellanic--type which are classified as irregulars with some spiral structure.  Coordinates, diameters, Galactic latitudes and types of galaxies are from {\it NED}.}
\tablenotetext{a}{\citet{Tully88} apart from NGC 4449 \citep{Summers03}}
\tablenotetext{b}{(Leiden/Argentine/Bonn) LAB Survey of Galactic H{\sc i}}
\tablenotetext{c}{Third Reference Catalog of Bright Galaxies}
\tablenotetext{d}{Physical scales corresponding to 0.5$\arcsec$ at the distance of each galaxy}
\tablenotetext{e}{\citet{Freedman88}}
\tablenotetext{f}{\citet{Saha94}}
\tablenotetext{g}{\citet{Annibali08}}
\tablenotetext{h}{\citet{Freedman94}}
\tablenotetext{i}{\citet{Tully88}}
\end{deluxetable}

\begin{deluxetable}{ccccl}
\tabletypesize{\scriptsize}
\tablecaption{Comparison of SNRs detected in this study with previous studies}
\tablewidth{0pt}
\tablehead{
\colhead{Galaxy} & \colhead{Number of } & \colhead{Number of } & \colhead{Number of} & \colhead{Number of} \\
\colhead{} &\colhead{Optically Identified} &\colhead{candidate radio} &\colhead{X-ray Identified} &\colhead{X-ray selected SNRs} \\
\colhead{} &\colhead{SNRs} &\colhead{SNRs} &\colhead{SNRs} &\colhead{in this study}\tablenotemark{\ast}}
\startdata
NGC 2403 &35\tablenotemark{a} &3\tablenotemark{b} &1\tablenotemark{c} & 15 (0)\\
NGC 5204 &3\tablenotemark{d} &0 &0 &0\\
NGC 4395 &0 &1\tablenotemark{e} &0 & 2 (1)\\
NGC 4449 &1\tablenotemark{f} &8\tablenotemark{g} &2\tablenotemark{h} & 4 (0)\\
NGC 3077 &0 &1\tablenotemark{i} &3\tablenotemark{j} & 5 (0)\\
NGC 4214 &0 &7\tablenotemark{e} &0 & 11 (4) \\
\enddata
\tablenotetext{\ast}{The numbers in parenthesis are the candidate SNRs in relation to the total (see Section 4 for the classification of X--ray SNRs). \\
References: (a) \citet{Dodorico80, MFBL97}; (b) \citet{TH94, Eck02}; (c) \citet{SP03, P07}; (d) \citet{MF97}; (e) \citet{Vukotic05, CW09}; (f) \citet{BH78, KB80, Blair83}; (g) \citet{Lacey07, CW09}; (h) \citet{VP97, PF03, Summers03}; (i) \citet{Rosa05}; (j) \citet{Ott03}}

\end{deluxetable}

\begin{deluxetable}{cccccc}
\tablecolumns {6}
\tablewidth{0pt}
\tablecaption{{\it Chandra} Observation Log}
\tablehead{
\colhead{Galaxy} & \colhead{Observation Number} & \colhead{Obs ID} & \colhead{Net Exposure}& \colhead{Date} & \colhead{Roll} \\
& & &\colhead{(ksec)} & &}
\startdata
\cutinhead{Single Exposures}
NGC 3077 & 1 & 2076 & 54.14 & 2001 Mar 07 & 199.5 \\
NGC 4449 & 1 & 2031 & 26.94 & 2001 Feb 04 & 108.8 \\
NGC 4395 & 1 & 882  & 17.19 & 2000 Jun 20 & 253.8 \\ 
\cutinhead{Multiple Exposures}
NGC 4214 & 1 & 2030 & 26.76 & 2001 Oct 16 & 23.9 \\
& 2 & 4743 & 27.56 & 2004 Apr 03 & 193.1 \\
& 3 & 5197 & 28.96 & 2004 Jul 30 & 275.8 \\
NGC 2403 & 1 & 2014 & 36.00 & 2001 Apr 17 & 266.8 \\
& 2 & 4628 & 47.12 & 2004 Aug 23 & 44.5 \\
& 3 & 4629 & 45.14 & 2004 Oct 03 & 75.2 \\
& 4 & 4630 & 50.58 & 2004 Dec 22 & 149.3 \\
NGC 5204 & 1 & 2028 & 10.15 & 2001 Jan 09 & 77.3 \\
& 2 & 2029 & 9.55 & 2001 May 02 & 197.7 \\
& 3 & 3933 & 48.94 & 2003 Aug 06 & 281.2 \\
& 4 & 3934 & 5.01 & 2003 Aug 09 & 284.4 \\
& 5 & 3935 & 4.76 & 2003 Aug 11 & 287.0 \\
& 6 & 3936 & 4.83 & 2003 Aug 14 & 290.0 \\
& 7 & 3937 & 4.91 & 2003 Aug 17 & 292.8 \\
& 8 & 3938 & 5.45 & 2003 Aug 19 & 295.7 \\
& 9 & 3939 & 5.44 & 2003 Aug 27 &304.2 \\
& 10 & 3940 & 5.13 & 2003 Sep 05 & 314.3 \\
& 11 & 3941 & 5.16 & 2003 Sep 14 & 324.2 \\
& 12 & 3942 & 5.51 & 2003 Sep 23 & 334.0 \\
& 13 & 3943 & 5.18 & 2003 Oct 03 & 345.5 \\
\enddata
\end{deluxetable}

\begin{deluxetable}{cc}
\tablecolumns {2}
\tablecaption{Number of background sources for the galaxies in our sample}
\tablewidth{0pt}
\tablehead{
\colhead{Galaxy} & \colhead{Expected background sources}
}
\startdata
NGC 2403 &60   \\
NGC 5204 &3    \\
NGC 4395 &0    \\
NGC 4449 &7.6  \\
NGC 3077 &6    \\
NGC 4214 &15.6 \\
\enddata
\tablecomments{Based on the 0.5 - 2.0 keV logN - logS of \citet{Kim07}, we estimated the number of background sources within the area of each galaxy covered by the {\it Chandra} observations. The {\it Chandra} data cover the entire $D_{25}$ area of NGC 4214 and NGC 4449, 30$\%$ of NGC 4395 and 80 - 90$\%$ of NGC 2403, NGC 3077 and NGC 5204.}
\end{deluxetable}

\begin{deluxetable}{ccccccccccccc}
\tablecolumns {13}
\rotate
\tabletypesize{\scriptsize}
\tablecaption{Photometric properties of potential SNRs in  NGC 3077}
\tablewidth{0pt}
\tablehead{
\multicolumn{3}{c}{} &\multicolumn{2}{c}{Soft Band} & \multicolumn{2}{c}{Medium Band} & \multicolumn{2}{c}{Hard Band} & \multicolumn{2}{c}{Total Band}\\
\cline{4-11}\\
\colhead{Src$\_$ID} & \colhead{RA} & \colhead{DEC} & \colhead{Obs} & \colhead{Bkg} &\colhead{Obs} & \colhead{Bkg} & \colhead{Obs} & \colhead{Bkg} & \colhead{Obs} & \colhead{Bkg} & \colhead{Area} & \colhead{Effective}\\
\colhead{} & \colhead{(h:m:s)} & \colhead{(d:m:s)} & \colhead{Counts} & \colhead{Counts} & \colhead{Counts} & \colhead{Counts} & \colhead{Counts} & \colhead{Counts}   & \colhead{Counts}  & \colhead{Counts}& \colhead{} & \colhead{Area}}
\startdata
LZB$\_$2  & 10:03:17.73 &+68:44:16.57 &  35 & 202  &   0  &  77 &   0 &   30 &  35 & 309 &  159.1 & 1.01 \\
LZB$\_$6  & 10:03:18.08 &+68:44:04.04 &  17 & 295  &   6  & 129 &   0 &   38 &  23 & 462 &  127.7 & 1.02 \\
LZB$\_$8  & 10:03:21.19 &+68:46:33.27 &   3 &  72  &   3  &  55 &   0 &   89 &   6 & 216 &  1035. & 0.99 \\
LZB$\_$12 & 10:03:21.82 &+68:45:03.29 &   5 &  88  &   7  &  53 &   0 &   76 &  12 & 217 &  424.6 & 0.99 \\
LZB$\_$13 & 10:03:12.14 &+68:43:19.07 &   1 &  65  &   7  &  74 &   2 &   70 &  10 & 209 &  652.5 & 0.99 \\
LZB$\_$14 & 10:03:08.25 &+68:44:08.57 &   8 &  47  &  15  &  44 &   5 &   53 &  28 & 144 &  395.0 & 0.99 \\
LZB$\_$15 & 10:03:20.60 &+68:41:40.29 &  75 &  49  & 109  &  60 &  23 &   73 & 207 & 182 &  213.0 & 0.99 \\
LZB$\_$18 & 10:03:18.62 &+68:43:56.87 &  29 & 295  &  86  & 129 &  11 &   38 & 126 & 462 &  182.2 & 1.02 \\
\enddata
\tablecomments{Column 1: Source ID; Columns 2-3: Source coordinates in J2000; Cols 4-11: Photometric counts of each source and its backgound in the soft, medium, hard and total bands respectively (see Section 3.3); Column 12: (Background / source) area ratio; Column 13: Effective area ratio with respect to a reference point (see \S3.3) in the soft energy band.}
\end{deluxetable}

\begin{deluxetable}{ccccccccccccc}
\tablecolumns {13}
\rotate
\tabletypesize{\scriptsize}
\tablecaption{Photometric properties of potential SNRs in NGC 4395}
\tablewidth{0pt}
\tablehead{
\multicolumn{3}{c}{} &\multicolumn{2}{c}{Soft Band} & \multicolumn{2}{c}{Medium Band} & \multicolumn{2}{c}{Hard Band} & \multicolumn{2}{c}{Total Band}\\
\cline{4-13}\\
\colhead{Src$\_$ID} & \colhead{RA} & \colhead{DEC} & \colhead{Obs} & \colhead{Bkg} &\colhead{Obs} & \colhead{Bkg} & \colhead{Obs} & \colhead{Bkg} & \colhead{Obs} & \colhead{Bkg} & \colhead{Area} & \colhead{Effective} \\
\colhead{} & \colhead{(h:m:s)} & \colhead{(d:m:s)} & \colhead{Counts} & \colhead{Counts} & \colhead{Counts} & \colhead{Counts} & \colhead{Counts} & \colhead{Counts}   & \colhead{Counts}  & \colhead{Counts} & \colhead{} & \colhead{Area}}
\startdata
LZB$\_$1  &12:25:54.52 &+33:30:44.29 &  16  & 62  &   2  & 27 &   0  &  32 &  18 & 121  & 597.5  &  1.00 \\
LZB$\_$2  &12:25:45.15 &+33:31:02.51 &  47  & 62  &   0  & 22 &   0  &  37 &  47 & 121  & 409.5  &  0.99 \\
LZB$\_$5  &12:25:47.01 &+33:36:06.22 &  21  & 59  &  14  & 29 &   2  &  65 &  37 & 153  & 232.5  &  0.99 \\
LZB$\_$6  &12:25:40.90 &+33:31:09.50 &   3  & 40  &   4  & 20 &   1  &  26 &   8 &  86  & 531.9  &  0.97 \\
LZB$\_$7  &12:25:55.03 &+33:30:15.02 &   1  & 42  &   3  & 23 &   0  &  21 &   4 &  86  & 1354.  &  1.25 \\
LZB$\_$10 &12:25:53.22 &+33:38:30.37 &   1  & 41  &   1  & 24 &   0  &  26 &   2 &  91  & 2492.  &  0.99 \\
LZB$\_$12 &12:25:59.77 &+33:33:20.57 &  35  &131  &  39  &123 &   8  &  39 &  82 & 293  & 292.6  &  1.76 \\
LZB$\_$14 &12:25:39.42 &+33:32:03.55 &  95  & 43  &  78  & 40 &   5  &  34 & 178 & 117  & 326.8  &  1.00 \\
LZB$\_$15 &12:25:48.93 &+33:32:00.81 &  85  & 58  & 114  & 41 &  48  &  60 & 247 & 159  & 471.4  &  0.99 \\
\enddata
\tablecomments{For the description of the columns see Table 5.}
\end{deluxetable}

\begin{deluxetable}{ccccccccccccc}
\tablecolumns {13}
\rotate
\tabletypesize{\scriptsize}
\tablecaption{Photometric properties of potential SNRs in NGC 4449}
\tablewidth{0pt}
\tablehead{
\multicolumn{3}{c}{} &\multicolumn{2}{c}{Soft Band} & \multicolumn{2}{c}{Medium Band} & \multicolumn{2}{c}{Hard Band} & \multicolumn{2}{c}{Total Band}\\
\cline{4-13}\\
\colhead{Src$\_$ID} & \colhead{RA} & \colhead{DEC} & \colhead{Obs} & \colhead{Bkg} &\colhead{Obs} & \colhead{Bkg} & \colhead{Obs} & \colhead{Bkg} & \colhead{Obs} & \colhead{Bkg} & \colhead{Area} & \colhead{Effective}\\
\colhead{} & \colhead{(h:m:s)} & \colhead{(d:m:s)} & \colhead{Counts} & \colhead{Counts} & \colhead{Counts} & \colhead{Counts} & \colhead{Counts} & \colhead{Counts}   & \colhead{Counts}  & \colhead{Counts}& \colhead{} & \colhead{Area}}
\startdata
LZB$\_$2  & 12:28:07.35 &+44:04:53.40 &  94 &    91 &  50 &  26  &  9 &   18 &  153 &  135  &   74.4 &  0.98 \\
LZB$\_$4  & 12:28:11.93 &+44:06:40.86 &  74 &   281 &  94 &  93  & 17 &   23 &  185 &  397  &  148.9 &  0.99 \\
LZB$\_$5  & 12:28:11.18 &+44:06:37.67 &  66 &   281 &  40 &  93  &  6 &   23 &  112 &  397  &   94.3 &  0.99 \\
LZB$\_$8  & 12:28:13.23 &+44:06:55.27 & 100 &   260 & 135 & 100  & 17 &   20 &  252 &  380  &  109.9 &  0.97 \\
LZB$\_$9  & 12:28:11.99 &+44:05:57.74 &  37 &   165 &  20 &  47  &  2 &   14 &   59 &  226  &   71.9 &  1.00  \\
LZB$\_$11 & 12:28:10.94 &+44:04:59.06 &   3 &   181 &  16 &  46  &  5 &   17 &   24 &  244  &  275.1 &  1.45  \\
LZB$\_$12 & 12:28:10.97 &+44:06:47.80 & 378 &   281 & 568 &  93  & 86 &   23 & 1032 &  397  &   82.4 &  0.99 \\
LZB$\_$13 & 12:28:10.93 &+44:03:37.58 & 403 &    74 &   2 &  37  &  0 &   50 &  405 &  161  &  107.6 &  0.99 \\
\enddata
\tablecomments{For the description of the columns see Table 5.}
\end{deluxetable}

\begin{deluxetable}{ccccccccccccc}
\tablecolumns {13}
\rotate
\tabletypesize{\scriptsize}
\tablecaption{Photometric properties of potential SNRs in NGC 4214}
\tablewidth{0pt}
\tablehead{
\multicolumn{3}{c}{} &\multicolumn{2}{c}{Soft Band} & \multicolumn{2}{c}{Medium Band} & \multicolumn{2}{c}{Hard Band} & \multicolumn{2}{c}{Total Band}\\
\cline{4-13}\\
\colhead{Src$\_$ID} & \colhead{RA} & \colhead{DEC} & \colhead{Obs} & \colhead{Bkg} &\colhead{Obs} & \colhead{Bkg} & \colhead{Obs} & \colhead{Bkg} & \colhead{Obs} & \colhead{Bkg} & \colhead{Area} & \colhead{Effective}\\
\colhead{} & \colhead{(h:m:s)} & \colhead{(d:m:s)} & \colhead{Counts} & \colhead{Counts} & \colhead{Counts} & \colhead{Counts} & \colhead{Counts} & \colhead{Counts}   & \colhead{Counts}  & \colhead{Counts}& \colhead{} & \colhead{Area}}
\startdata
LZB$\_$5  &12:15:40.82  &+36:19:26.29 &  10  &   2490  &    2   &  1534  &    4   &   1861 &  16 & 5885  & 4939.6       &   1.00  \\
LZB$\_$7  &12:15:33.41  &+36:18:58.88 &  50  &    925  &    5   &   809  &    3   &   1229 &  58 & 2963  & 1158.3       &   0.98  \\
LZB$\_$10 &12:15:49.71  &+36:18:46.69 &   3  &    924  &    5   &   788  &    1   &   1177 &   9 & 2889  & 3948.4       &   0.99  \\
LZB$\_$11 &12:15:37.93  &+36:22:21.00 &   2  &    867  &    5   &   884  &    0   &   1091 &   7 & 2842  & 4956.7       &   0.98  \\
LZB$\_$16 &12:15:40.16  &+36:19:25.21 &   4  &   2490  &    3   &  1534  &    0   &   1861 &   7 & 5885  & 6656.5       &   1.05  \\
LZB$\_$20 &12:15:38.13  &+36:20:50.08 &   6  &   1047  &    19  &  955   &    5   &   1335 &  30 & 3337  & 3237.0       &   1.04  \\ 
LZB$\_$23 &12:15:48.80  &+36:17:01.83 &   2  &    834  &    8   &   883  &    1   &   1437 &  11 & 3154  & 4187.0       &   0.96  \\
LZB$\_$24 &12:15:41.37  &+36:21:13.67 & 135  &   1047  &  235   &   955  &  110   &   1335 & 480 & 3337  & 435.43       &   0.98  \\
\enddata
\tablecomments{For the description of the columns see Table 5.}
\end{deluxetable}

\begin{deluxetable}{ccccccccccccc}
\tablecolumns {13}
\rotate
\tabletypesize{\scriptsize}
\tablecaption{Photometric properties of potential SNRs in NGC 2403}
\tablewidth{0pt}
\tablehead{
\multicolumn{3}{c}{} &\multicolumn{2}{c}{Soft Band} & \multicolumn{2}{c}{Medium Band} & \multicolumn{2}{c}{Hard Band} & \multicolumn{2}{c}{Total Band}\\
\cline{4-13}\\
\colhead{Src$\_$ID} & \colhead{RA} & \colhead{DEC} & \colhead{Obs} & \colhead{Bkg} &\colhead{Obs} & \colhead{Bkg} & \colhead{Obs} & \colhead{Bkg} & \colhead{Obs} & \colhead{Bkg} & \colhead{Area} & \colhead{Effective}\\
\colhead{} & \colhead{(h:m:s)} & \colhead{(d:m:s)} & \colhead{Counts} & \colhead{Counts} & \colhead{Counts} & \colhead{Counts} & \colhead{Counts} & \colhead{Counts}   & \colhead{Counts}  & \colhead{Counts}& \colhead{} & \colhead{Area}}
\startdata
LZB$\_$2   &07:37:10.70 &+65:33:11.73 &  28  &    1816  &    49  &    1598  &    10  &       2340 &   87   &      5754&   781.4 &   0.95  \\
LZB$\_$5   &07:36:52.44 &+65:36:41.69 &  34  &    4482  &     7  &    2854  &     2  &       3418 &   43   &     10754&   2574. &   0.97  \\
LZB$\_$14  &07:37:16.43 &+65:33:29.24 &  28  &    1818  &    12  &    1609  &     1  &       2346 &   41   &      5773&   1090. &   0.98  \\
LZB$\_$21  &07:38:00.72 &+65:32:09.30 &  81  &     519  &   148  &     491  &    42  &        785 &  271   &      1795&   241.7 &   1.18  \\
LZB$\_$30  &07:36:46.13 &+65:36:41.80 &  32  &    4482  &    17  &    2854  &     1  &       3418 &   50   &     10754&   1035. &   0.95  \\
LZB$\_$39  &07:36:35.95 &+65:36:09.35 &  27  &    4482  &    56  &    2854  &    20  &       3418 &  103   &     10754&   1368. &   1.18  \\
LZB$\_$41  &07:37:13.50 &+65:35:58.72 &  49  &    3543  &    12  &    2389  &     4  &       3112 &   65   &      9044&   2123. &   0.98  \\
LZB$\_$42  &07:36:57.54 &+65:36:04.45 &  37  &    4482  &    24  &    2854  &     2  &       3418 &   63   &     10754&   2149. &   0.91  \\
\enddata
\tablecomments{For the description of the columns see Table 5.}
\end{deluxetable}

\begin{deluxetable}{ccccccccccccc}
\tablecolumns {13}
\rotate
\tabletypesize{\scriptsize}
\tablecaption{Photometric properties of potential SNRs in NGC 5204}
\tablewidth{0pt}
\tablehead{
\multicolumn{3}{c}{} &\multicolumn{2}{c}{Soft Band} & \multicolumn{2}{c}{Medium Band} & \multicolumn{2}{c}{Hard Band} & \multicolumn{2}{c}{Total Band}\\
\cline{4-13}\\
\colhead{Src$\_$ID} & \colhead{RA} & \colhead{DEC} & \colhead{Obs} & \colhead{Bkg} &\colhead{Obs} & \colhead{Bkg} & \colhead{Obs} & \colhead{Bkg} & \colhead{Obs} & \colhead{Bkg} & \colhead{Area} & \colhead{Effective}\\
\colhead{} & \colhead{(h:m:s)} & \colhead{(d:m:s)} & \colhead{Counts} & \colhead{Counts} & \colhead{Counts} & \colhead{Counts} & \colhead{Counts} & \colhead{Counts}   & \colhead{Counts}  & \colhead{Counts}& \colhead{} & \colhead{Area}}
\startdata  
LZB$\_$8  &13:29:39.26 &+58:25:31.20 &    94 &   834 &      47 &   595 &       9  &  586&   150&  2015& 235.7 &  0.98\\
LZB$\_$10 &13:29:38.63 &+58:25:05.51 & 14427 &  1182 &   13451 &   834 &    2191  &  846& 30069&  2862&  71.6 &  0.99\\
LZB$\_$11 &13:29:27.47 &+58:25:34.26 &    86 &    125&     124 &   144 &      56  &  275&   266&   544&  91.9 &  0.99\\ 
\enddata
\tablecomments{For the description of the columns see Table 5.}
\end{deluxetable}

\begin{deluxetable}{ccccccc}
\tablecolumns {7}
\tabletypesize{\scriptsize}
\tablecaption{X--ray colors for potential SNRs in NGC 3077}
\tablewidth{0pt}
\tablehead{
\colhead{Source$\_$ID}  & \colhead{Col(S/M)} & \colhead{Col(M/H)}& \colhead{Col(S/H)} &\colhead{Spectra} &\colhead{Classification} &\colhead{Counterpart}}
\startdata
LZB$\_$2   &  $1.84^{+0.95}_{-0.57}$    &  $0.00^{+1.18}_{-1.18}$ &   $1.84^{+0.95}_{-0.57}$  &Yes &SSS           & SSS (Ott-S4)\tablenotemark{a}\\
LZB$\_$6   &  $0.44^{+0.26}_{-0.23}$   &  $1.03^{+0.98}_{-0.61}$ &   $1.48^{+0.95}_{-0.58}$   &Yes &probable SNR  &SNR (Ott-S6)\tablenotemark{a}\\
LZB$\_$8   &  $0.00^{+0.35}_{-0.35}$    &  $0.84^{+0.98}_{-0.64}$ &   $0.84^{+0.98}_{-0.64}$  &Yes & XRB          &\nodata \\
LZB$\_$12  &  $-0.14^{+0.25}_{-0.26}$  &  $1.16^{+0.96}_{-0.58}$ &   $1.03^{+0.96}_{-0.61}$  &Yes & probable SNR &\nodata\\
LZB$\_$13  &  $-0.72^{+0.41}_{-0.54}$  &  $0.49^{+0.38}_{-0.33}$ &   $-0.23^{+0.54}_{-0.60}$ &Yes & probable SNR &\nodata\\
LZB$\_$14  &  $-0.26^{+0.18}_{-0.19}$  &  $0.46^{+0.23}_{-0.21}$ &   $0.19^{+0.25}_{-0.24}$  &Yes & XRB          & \nodata\\
LZB$\_$15  &  $-0.16^{+0.09}_{-0.22}$  &  $0.68^{+0.10}_{-0.10}$ &   $0.51^{+0.10}_{-0.10}$  &Yes & probable SNR & \nodata\\
LZB$\_$18  &  $-0.49^{+0.09}_{-0.10}$  &  $0.88^{+0.15}_{-0.13}$ &   $0.39^{+0.16}_{-0.15}$  &Yes & probable SNR &SNR (Ott-S1)\tablenotemark{a,}\tablenotemark{b} \\
\enddata
\tablecomments{Column 1: Source ID; Columns 2-4: X--ray colors and their errors (derived from the BEHR method) in the Soft/Medium, Medium/Hard and Soft/Hard bands respectively; Column 5: Whether spectrum of the source is derived; Column 6: The classification of the source according to its spectrum (see \S3.4.2 and \S4) and Column 7: Multiwavelength associations}
\tablenotetext{a}{Sources identified by \citet{Ott03}.}
\tablenotetext{b}{Radio candidate SNR by \citet{Rosa05}}
\end{deluxetable}

\begin{deluxetable}{ccccccc}
\tablecolumns {7}
\tabletypesize{\scriptsize}
\tablecaption{X--ray colors for potential SNRs in NGC 4395}
\tablewidth{0pt}
\tablehead{
\colhead{Source$\_$ID} & \colhead{Col(S/M)} & \colhead{Col(M/H)}& \colhead{Col(S/H)} &\colhead{Spectra} &\colhead{Classification} &\colhead{Counterpart}}
\startdata
LZB$\_$1   &   $0.82^{+ 0.34 }_{-0.28 }$   &  $ 0.69 ^{+0.99 }_{-0.67 }$ &  $ 1.52 ^{+0.95 }_{-0.57 }$ &Yes  & XRB  &\nodata\\
LZB$\_$2   &   $1.97^{+ 0.95 }_{-0.55 }$   &  $ 0.00 ^{+1.18 }_{-1.18 }$ &  $ 1.97 ^{+0.95 }_{-0.55 }$ &Yes  & XRB  &\nodata\\
LZB$\_$5   &   $0.17^{+ 0.15 }_{-0.15 }$   &  $ 0.80 ^{+0.42 }_{-0.32 }$ &  $ 0.97 ^{+0.41 }_{-0.32 }$ &Yes  & XRB  &\nodata\\
LZB$\_$6   &   $-0.11^{+ 0.32 }_{-0.34 }$  &  $ 0.49 ^{+0.52 }_{-0.43 }$ &  $ 0.37 ^{+0.54 }_{-0.44 }$ &Yes  & XRB  &\nodata\\
LZB$\_$7   &   $-0.37^{+ 0.43 }_{-0.52 }$  &  $ 0.84 ^{+0.98 }_{-0.63 }$ &  $ 0.46 ^{+1.04 }_{-0.77 }$ &Yes  & XRB  &\nodata\\
LZB$\_$10  &   $0.00^{+ 0.57 }_{-0.58 }$  &  $ 0.47 ^{+1.03 }_{-0.75 }$ &  $ 0.47 ^{+1.03 }_{-0.77 }$ &No   & candidate SNR\tablenotemark{a}&\nodata\\
LZB$\_$12  &   $-0.05^{+ 0.10 }_{-0.10 }$  &  $ 0.67 ^{+0.17 }_{-0.16 }$ &  $ 0.62 ^{+0.18 }_{-0.16 }$ &Yes  & XRB  &\nodata\\
LZB$\_$14  &   $0.09^{+ 0.06 }_{-0.07 }$  &  $1.16^{+ 0.21 }_{-0.18 }$  &  $ 1.25 ^{+0.21 }_{-0.18 }$ &Yes  & SNR  &\nodata\\
LZB$\_$15  &   $-0.13^{+ 0.06 }_{-0.06 }$    &$0.37^{+ 0.08 }_{-0.06 }$   &$-0.13^{+ 0.06 }_{-0.06 }$   &Yes  & XRB &\nodata\\ 
\enddata
\tablecomments{For the description of the Columns see Table 11.}
\tablenotetext{a}{We do not have spectrum for this source (due to small number of counts) but its large errobars at the color--color diagram place it on the region of candidate SNRs}.
\end{deluxetable}

\begin{deluxetable}{ccccccc}
\tablecolumns {7}
\tabletypesize{\scriptsize}
\tablecaption{X--ray colors for potential SNRs in NGC 4449}
\tablewidth{0pt}
\tablehead{
\colhead{Source$\_$ID} & \colhead{Col(S/M)} & \colhead{Col(M/H)}& \colhead{Col(S/H)} &\colhead{Spectra} &\colhead{Classification} &\colhead{Counterpart}\tablenotemark{a}}
\startdata
LZB$\_$2    &   $0.27  ^{+ 0.08 }_{-0.08 }$  &  $0.74 ^{+ 0.16 }_{-0.16}$ &  $1.00   ^{+0.16}_{-0.15 }$ &Yes  & XRB          &SNR/XRB                 \\
LZB$\_$4    &   $-0.11  ^{+ 0.07 }_{-0.07 }$ &  $0.73 ^{+ 0.12 }_{-0.11}$ &  $0.62   ^{+0.12}_{-0.11 }$ &Yes  & XRB          &SNR                     \\
LZB$\_$5    &   $0.19 ^{+ 0.1 }_{-0.78 }$    &  $0.80 ^{+ 0.20 }_{-0.18}$ &  $1.00   ^{+0.20}_{-0.18 }$ &Yes  & XRB          &SNR/XRB                 \\
LZB$\_$8    &   $-0.14  ^{+ 0.05 }_{-0.06 }$ &  $0.89 ^{+ 0.12 }_{-0.11}$ &  $0.75   ^{+0.12}_{-0.11 }$ &Yes  & XRB          &SNR/XRB                 \\
LZB$\_$9    &   $ 0.25 ^{+ 0.13 }_{-0.12 }$  &  $0.93 ^{+ 0.39 }_{-0.31}$ &  $1.18   ^{+0.38}_{-0.30 }$ &Yes  & SNR          &SNR/SSS                 \\
LZB$\_$11   &   $-0.75 ^{+ 0.31 }_{-0.39 }$  &  $0.48 ^{+ 0.23 }_{-0.21}$ &  $-0.27  ^{+0.36}_{-0.42 }$ &Yes  & XRB          &XRB                     \\
LZB$\_$12   &   $-0.18  ^{+ 0.03 }_{-0.03 }$ &  $0.82 ^{+ 0.05 }_{-0.05}$ &  $0.64^{+ 0.03 }_{-0.05}$   &Yes  & SNR          &SNR \tablenotemark{b}   \\
LZB$\_$13   &   $2.25  ^{+ 0.43 }_{-0.31 }$  &  $0.63 ^{+ 1.06 }_{-0.77}$ &  $2.89   ^{+0.95}_{-0.55 }$ &Yes  & SSS          &SSS                     \\
\enddata
\tablecomments{For the description of the Columns see Table 11.}
\tablenotetext{a}{X-ray detected sources by \citet{Summers03}}
\tablenotetext{b}{The most oxygen rich SNR known (see $\S$2.1)}
\tablenotetext{c}{It is also detected in radio as H{\sc ii} region by \citet{CW09}}
 \end{deluxetable}

\begin{deluxetable}{ccccccc}
\tablecolumns {7}
\tabletypesize{\scriptsize}
\tablecaption{X--ray colors for potential SNRs in NGC 4214}
\tablewidth{0pt}
\tablehead{
\colhead{Source$\_$ID} & \colhead{Col(S/M)} & \colhead{Col(M/H)}& \colhead{Col(S/H)} & \colhead{Spectra} &\colhead{Classification} &\colhead{Counterpart}}
\startdata
LZB$\_$5   &   $0.65 ^{+0.44}_{-0.35 }$  &  $-0.27^{+0.42 }_{-0.48}$ &  $0.38 ^{+0.29 }_{-0.26}$  &Yes & XRB          &\nodata  \\
LZB$\_$7   &   $1.01 ^{+0.25}_{-0.21 }$  &  $0.27 ^{+0.55 }_{-0.44}$ &  $1.28 ^{+0.50 }_{-0.33}$  &Yes & SNR          &\nodata  \\
LZB$\_$10  &   $-0.21^{+0.32}_{-0.35 }$  &  $0.61 ^{+0.69 }_{-0.49}$ &  $0.40 ^{+0.73 }_{-0.54}$  &No  & candidate SNR&Background galaxy \tablenotemark{d}  \\
LZB$\_$11  &   $-0.35^{+0.36}_{-0.41 }$  &  $1.03 ^{+0.98}_{-0.61 }$ &  $0.66 ^{+1.03}_{-0.70 }$  &No  & candidate SNR     &\nodata  \\
LZB$\_$16  &   $0.10 ^{+0.37}_{-0.36 }$  &  $0.81 ^{+0.99 }_{-0.64}$ &  $0.92 ^{+0.98 }_{-0.63}$  &No  & candidate SNR&\nodata  \\ 
LZB$\_$20  &   $-0.49^{+0.20}_{-0.22 }$  &  $0.57 ^{+0.25}_{-0.21 }$ &  $0.09 ^{+0.28}_{-0.28 }$  &Yes  & XRB       &\nodata  \\
LZB$\_$23  &   $-0.55^{+0.34}_{-0.40 }$  &  $0.81 ^{+0.70 }_{-0.46}$ &  $0.26^{+0.80 }_{-0.61}$  &No  & candidate SNR&\nodata  \\  
LZB$\_$24  &   $-0.24^{+0.05}_{-0.05 }$  &  $0.34^{+0.05}_{-0.0.05}$    &$0.08^{+0.07}_{-0.04}$   &Yes & XRB          &H{\sc ii}\tablenotemark{c} \\
\enddata
\tablecomments{For the description of the Columns see Table 11. Sources 10, 11, 16 and 23 do not have adequate number of counts to allow us the extraction of their spectra, but their X--ray colors (within their errobars) place them in the region of candidate SNRs.}
\tablenotetext{a}{Radio candidate SNR by \citet{Vukotic05}, the source is denoted as ''$\varrho$''.}
\tablenotetext{b}{Radio candidate SNRs by \citet{CW09}}
\tablenotetext{c}{Radio classified  H{\sc ii} regions by \citet{CW09}}
\tablenotetext{d}{The classification criteria of \citet{CW09} for radio SNRs is the radio spectral index to be $\alpha$$\leq$-0.2 and H$\alpha$ emission associated with the source. In the case of this source, the radio spectral index is -0.32 but there is no detected H$\alpha$ emission, therefore they classify this source as a radio background galaxy.}
\end{deluxetable}

\begin{deluxetable}{ccccccc}
\tablecolumns {7}
\tabletypesize{\scriptsize}
\tablecaption{X--ray colors for potential SNRs in NGC 2403}
\tablewidth{0pt}
\tablehead{
\colhead{Source$\_$ID} & \colhead{Col(S/M)} & \colhead{Col(M/H)}& \colhead{Col(S/H)} &\colhead{Spectra} &\colhead{Classification} &\colhead{Counterpart}}
\startdata
LZB$\_$2    &   $-0.26 ^{+0.11 }_{-0.11 }$   &  $0.79 ^{+0.23 }_{-0.19 }$  &  $0.54^{+0.24 }_{-0.20 }$   &Yes & probable SNR & SNR (MFBL--29)\tablenotemark{a}  \\ 
LZB$\_$5    &   $0.70  ^{+0.23 }_{-0.19 }$   &  $0.61 ^{+0.81 }_{-0.49 }$  &  $1.32^{+0.80 }_{-0.41 }$   &Yes &probable SNR &\nodata  \\
LZB$\_$14   &   $0.38  ^{+0.17 }_{-0.16 }$   &  $1.16 ^{+0.95 }_{-0.54 }$  &  $1.55^{+0.95 }_{-0.52 }$   &Yes & probable SNR & SNR (MFBL-31,SP-30)\tablenotemark{b}  \\
LZB$\_$21   &   $-0.26 ^{+0.06 }_{-0.06 }$   &  $0.55 ^{+0.11 }_{-0.05 }$  &  $0.30^{+0.09 }_{-0.08 }$   &Yes & XRB    &\nodata \\  
LZB$\_$30   &   $0.28  ^{+0.16 }_{-0.15 }$   &  $1.35 ^{+0.95 }_{-0.55 }$  &  $1.62^{+0.95 }_{-0.54 }$   &Yes & probable SNR & SNR (MFBL-7, Eck-$\mu$ ?) \\
LZB$\_$39   &   $-0.35 ^{+0.11 }_{-0.11 }$   &  $0.48 ^{+0.13 }_{-0.12 }$  &  $0.13^{+0.14 }_{-0.14 }$   &Yes &XRB      &\nodata \\
LZB$\_$41   &   $0.62  ^{+0.16 }_{-0.14 }$   &  $0.55 ^{+0.47 }_{-0.34 }$  &  $1.17^{+0.45 }_{-0.30 }$   &Yes & probable SNR &\nodata \\
LZB$\_$42   &   $0.18  ^{+0.12 }_{-0.12 }$   &  $1.21 ^{+0.83 }_{-0.44 }$  &  $1.39^{+0.83 }_{-0.44 }$   &Yes & SNR    & SNR (MFBL-17)\tablenotemark{c} \\
\enddata
\tablecomments{For the description of the Columns see Table 11.}
\tablecomments{(MFBL): Optically identified SNRs by \citet{MFBL97}; (SP): X--ray identified SNRs by \citet{SP03}; (TH): Radio candidate SNR by \citet{TH94}; (Eck): Radio candidate SNR by \citet{Eck02}. The questionmarks correspond to large position offsets. }
\tablenotetext{a}{This source is also X--ray detected (SP--31) by \citet{SP03}.} 
\tablenotetext{b}{Also X--ray detected by \citet{P07}.} 
\tablenotetext{c}{This source is also X--ray detected (SP--16) by \citet{SP03} and \citet{P07}.} 
\tablenotetext{d}{XRB by \citet{SP03} while it is also detected as ULX by \citet{Winter06}}
\tablenotetext{e}{\citet{SP03} have extracted spectra for this source (SP20) without defining any classification. Also detected as ULX by \citet{Winter06}}
\tablenotetext{f}{\citet{Winter06}.}
\tablenotetext{g}{Also X--ray detected (SP--12) by \citet{SP03} and \citet{P07}. }
\tablenotetext{h}{Classified as ULX by \citet{Winter06} and \citet{SP03}.}
\end{deluxetable}

\begin{deluxetable}{ccccccc}
\tablecolumns {7}
\tabletypesize{\scriptsize}
\tablecaption{X--ray colors for potential SNRs in NGC 5204}
\tablewidth{0pt}
\tablehead{
\colhead{Source$\_$ID} & \colhead{Col(S/M)} & \colhead{Col(M/H)}& \colhead{Col(S/H)} &\colhead{Spectra} &\colhead{Classification} &\colhead{Counterpart}}
\startdata
LZB$\_$8   &   $0.30 ^{+0.08 }_{-0.08 }$  & $0.80 ^{+0.20 }_{-0.20 }$ & $1.1 ^{+0.20 }_{-0.20 }$    &Yes  &XRB     &\nodata\\
LZB$\_$10  &   $0.03 ^{+0.006 }_{-0.005}$ & $0.79 ^{+0.01 }_{-0.01 }$ & $0.82 ^{+0.01 }_{-0.01 }$   &Yes  &XRB     &\nodata\\
LZB$\_$11  &   $-0.16 ^{+0.06 }_{-0.06}$  & $0.36 ^{+0.07 }_{-0.07 }$ & $0.22 ^{+0.07 }_{-0.10 }$   &Yes  &XRB     &\nodata\\
\enddata
\tablecomments{For the description of the Columns see Table 11.}
\end{deluxetable}

\clearpage

\begin{deluxetable}{ccccccccccc}
\rotate
\tabletypesize{\scriptsize}
\tablecolumns {11}
\tablewidth{0pt}
\tablecaption{Spectral fitting results of the point-sources in NGC 3077}
\tablehead{
\colhead{Source$\_$ID} & \colhead{Model} & \colhead{N$_{\rm H}$} & \colhead{$\Gamma/$kT} & \colhead{Norm} & \colhead{$\chi^2$/dof} & \colhead{Goodness($\%$)} &\colhead{F$_{abs}$}  &\colhead{F$_{unabs}$} & \colhead{logL$_{X}^{unabs}$} &  \colhead{Classification}  \\
\colhead{} & \colhead{} & \colhead{($\times$ 10$^{22}$ $cm^{-2}$)} & \colhead{(Index/keV)}  & \colhead{($10^{-5}$ A)}  &\colhead{or Cstat}  &\colhead{(for Cstat)} &\multicolumn{2}{c}{($\times 10^{-15}$ erg s$^{-1} cm^{-1}$)} & \colhead{(erg s$^{-1}$)} & \colhead{}} 
\startdata
\cutinhead{chi-square statistics}
LZB$\_$15 & apec      &$0.06^{+0.05}_{-0.04}$  & $2.8^{+1.18}_{-0.68}$  & $1.5^{+0.25}_{-0.26}$   & 4.98$/$10   &\nodata & 21   & 24   & 37.6 & probable SNR  \\
LZB$\_$19 & PL &$0.14^{+0.13}_{-0.12}$  & $1.38^{+0.31}_{-0.19}$ & $0.32^{+0.21}_{-0.12}$         & 4.15/6      & ..     & 7.9  & 58   & 38.0 & XRB\\
\cutinhead{cash statistics}
LZB$\_$2 & bbody     &$0.41^{+0.14}_{-0.12}$  &$0.02^{+0.09}_{-0.01}$ &$>936.2$                   & 1.84        &56.0    & 3.2  & 5500 & 39.9 & SSS\tablenotemark{b}\\
LZB$\_$6 & apec      &$0.34^{+0.43}_{-0.34}$  &$0.30^{+0.45}_{-0.22}$ & $>0.44$                   & 0.56        &46.9    & 1.5  & 8.3  & 37.1 & probable SNR\tablenotemark{b}\\
LZB$\_$8 & PL &0.05\tablenotemark{a}   &$3.29^{+2.32}_{-2.09}$ &$0.02^{+0.02}_{-0.02}$            & 2.48        &20.5    & 0.8  & 1.2  & 36.3 & XRB\\
LZB$\_$12& apec      &$0.70^{+0.53}_{-0.70}$  &0.77 ($>0.18$)          &$0.18^{+2.8}_{-0.18}$     & 3.54        &46.6    & 0.8  & 4.6  & 36.8 & probable SNR\\
LZB$\_$13& apec      &$2.97^{+4.8}_{-1.4}$    &$0.19^{+0.43}_{-0.19}$  &$>111.1$                  & 1.44        &78.3    & 0.7  & 1800 & 39.4 & probable SNR\\
LZB$\_$14& PL & 0.05\tablenotemark{a}  &$1.35^{+0.76}_{-0.76}$ &$0.06^{+0.03}_{-0.002}$           & 2.22        &11.1    & 5.9  & 6.3  & 36.8 & XRB\\
\enddata
\tablecomments{Column 1: Source ID; Column 2: Fitted model; Column 3: H{\sc i} Column Density; Column 4: Best fit value of the power law photon index or the thermal component temperature in keV; Column 5: The normalization factor given in units of $10^{-5}$ K. For apec A is: $10^{-14} (4 \pi D_{A}^2)^{-1} \int n_{e} n_{p}$dV, where $D_{A}$ is the angular distance to the source in centimeters and $n_{e} n_{p}$ are the electron and proton densities respectively in $cm^{-2}$, for PL: photons keV$^{-1}$ cm$^{-2}$ at 1 keV and for bbody: 10$^{39}$ erg s$^{-1}$ (10 kpc)$^{-1}$; Column 6: $\chi^2$/degrees of freedom or Cash statistic; Column 7: Goodness-of fit for Cash statistics; Column 8: Absorbed flux in 0.3 - 10 keV band; Column 9: Unabsorbed flux in 0.3 - 10 keV band; Column 10: Source classification}
\tablenotetext{a}{Fixed to Galactic line-of-sight value}
\tablenotetext{b}{\citet{Ott03} have classified our LZB6, LZB18 as X--ray SNRs and LZB2 as a SSS}
\tablenotetext{c}{Radio candidate SNR by \citet{Rosa05}}
\end{deluxetable}

\begin{deluxetable}{ccccccccccc}
\rotate
\tabletypesize{\scriptsize}
\tablecolumns {11}
\tablewidth{0pt}
\tablecaption{Spectral fitting results of the point-sources in NGC 4395}
\tablehead{\colhead{Source} & \colhead{Model} & \colhead{N$_{\rm H}$} & \colhead{$\Gamma$/kT} & \colhead{Norm} & \colhead{$\chi^2$/dof} & \colhead{Goodness($\%$)} &\colhead{F$_{abs}$}  &\colhead{F$_{unabs}$} & \colhead{logL$_{X}^{unabs}$} & \colhead{Classification}\\
\colhead{} & \colhead{} & \colhead{($\times 10 ^{22} cm^{-2}$)} & \colhead{(Index/keV)}  & \colhead{($10^{-5}$ A)}  &\colhead{or Cstat}  &\colhead{(for Cstat)} & \multicolumn{2}{c}{($\times 10^{-14}$ erg s$^{-1} cm^{-1}$)} & \colhead{(erg s$^{-1}$)} & \colhead{}}
\startdata
\cutinhead{chi-square statistics}
LZB$\_$14 & apec + apec    &$0.545$                     &1.37                     &94                        &8.65/4    & \nodata  & 3.1     & 75.4    & 38.8  & SNR  \\
\nodata   &\nodata         &\nodata                     &0.17                     &10.1                      &\nodata   &\nodata   & \nodata & \nodata & 37.8  & \nodata\\
LZB$\_$15 & PL             &$0.03^{+0.06}_{-0.03}$      &$1.37^{+0.32}_{-0.25}$   &$1.27^{+0.48}_{-0.26}$    &16.25/13  &\nodata   & 12      &12       & 38.0  & XRB \\
\cutinhead{cash statistics}
LZB$\_$1 & PL  & 0.0185\tablenotemark{a}  &$0.64^{+1.46}_{-3.64}$   &$0.30^{+0.27}_{-0.15}$        &3.62 &60.0  & 7.9    & 8.0    & 37.8 &XRB\\
LZB$\_$2 & PL  &$0.03^{+0.19}_{-0.03}$    &$3.36^{+3.13}_{-1.70}$   &$0.33^{+0.46}_{-0.33}$        &2.74 &21.1  & 1.5    & 2.0    & 37.2 &XRB \\
LZB$\_$5 & PL  & 0.0185\tablenotemark{a}  &$2.14^{+0.59}_{-0.58}$   &$0.24^{+0.08}_{-0.06}$        &2.78 &7.3   & 1.2    & 1.3    & 37.0 &XRB\\
LZB$\_$6 & PL  & 0.0185\tablenotemark{a}  &$1.27^{+1.26}_{-1.26}$   &$0.05^{+0.04}_{-0.05}$        &9.75 &84.4  & 0.6    & 0.6    & 36.7 &XRB \\
LZB$\_$7 & PL  & 0.0185\tablenotemark{a}  &$2.06^{+4.30}_{-5.06}$   &$0.05^{+0.06}_{-0.05}$        &4.13 &74.2  & 0.3    & 0.3    & 36.4 &XRB \\
\enddata
\tablecomments{For the description of the Columns see Table 17.}
\tablenotetext{a}{Fixed to Galactic line-of-sight value}
\end{deluxetable}

\clearpage

\begin{deluxetable}{ccccccccccc}
\rotate
\tabletypesize{\scriptsize}
\tablecolumns {11}
\tablewidth{0pt}
\tablecaption{Spectral fitting results of the point-sources in NGC 4449}
\tablehead{\colhead{Source} & \colhead{Model} & \colhead{N$_{\rm H}$} & \colhead{$\Gamma$/kT} & \colhead{Norm} & \colhead{$\chi^2$/dof} & \colhead{Goodness($\%$)} &\colhead{F$_{abs}$}  &\colhead{F$_{unabs}$} & \colhead{logL$_{X}^{unabs}$} & \colhead{Classification\tablenotemark{a}}\\
\colhead{} & \colhead{} & \colhead{($\times$ 10$^{22}$ $cm^{-2}$)} & \colhead{(Index/keV)}  & \colhead{($10^{-5}$ A)}  &\colhead{or Cstat}  &\colhead{(for Cstat)} &\multicolumn{2}{c}{($\times 10^{-14}$ erg s$^{-1} cm^{-1}$)} & \colhead{(erg s$^{-1}$)} & \colhead{}}
\startdata
\cutinhead{chi-square statistics} 
LZB$\_$2    &PL         &$0.03^{+0.11}_{-0.03}$    &$2.46^{+1.37}_{-0.48}$           &$0.61^{+0.55}_{-0.14}$    & 7.43/6       &\nodata  & 2.5   &3.0    & 37.8 &XRB\\
\nodata     &\nodata    &\nodata                   &kT=$1.25^{+0.90}_{-0.32}$        &$>0.13$                   &\nodata       &\nodata  &\nodata &\nodata & \nodata &\nodata\\
LZB$\_$8    &PL         &$0.20^{+0.15}_{-0.09}$    &$2.48^{+0.49}_{-0.53}$           &$1.94^{+1.58}_{-0.82}$    &1.45/6        &\nodata  &5.0     &9.4    & 38.3 &XRB\\
LZB$\_$12   &PL + 2apec &$0.30^{+0.44}_{-0.09}$    &$\Gamma=1.67^{+0.21}_{-0.75}$    &$1.62^{+1.47}_{-1.46}$    &26.7/27       &\nodata  &18      &41      & 38.9 &SNR\tablenotemark{c}\\
\nodata     &\nodata    &\nodata                   &$kT_{1}= 0.25^{+0.09}_{-0.11}$   &$>2.80$                   &\nodata       &\nodata  &\nodata &\nodata & \nodata &\nodata\\
\nodata     &\nodata    &\nodata                   &$kT{_2}=1.27^{+0.09}_{-0.11}$    &$8.00^{+7.00}_{-2.50}$     &\nodata       &\nodata  &\nodata &\nodata & \nodata &\nodata\\ 
LZB$\_$13   &bbody      &$0.01^{+0.07}_{-0.01}$    &$0.09^{+0.01}_{-0.01}$           &$0.15^{+0.22}_{-0.04}$    &18.7/16       &\nodata  &6.3     &7.5    & 38.2 &SSS\\
LZB$\_$16   &PL         &$0.07^{+0.04}_{-0.03}$    &$2.13^{+0.24}_{-0.2}$            &$5.76^{+1.01}_{-0.79}$    &53.9/36       &\nodata  &24      &30      & 38.8 &XRB\\
\enddata
\tablecomments{For the description of the Columns see Table 17.}
\tablenotetext{a}{Already X--ray detected sources by \citet{Summers03}}
\tablenotetext{b}{Fixed to Galactic line-of-sight value}
\tablenotetext{c}{Oxygen rich SNR detected in all bands}
\tablenotetext{d}{Radio observations by \citet{CW09} denote this source as an H{\sc ii} region.}
\end{deluxetable}

\clearpage

\begin{deluxetable}{rcccccccccc}
\rotate
\tabletypesize{\scriptsize}
\tablecolumns {11}
\tablecaption{Spectral fitting results of the point-sources in NGC 4214}
\tablewidth{0pt}
\tablehead{\colhead{Source} & \colhead{Model} & \colhead{N$_{\rm H}$} & \colhead{$\Gamma$/kT} & \colhead{Norm} & \colhead{$\chi^2$/dof} & \colhead{Goodness($\%$)} &\colhead{F$_{abs}$}  &\colhead{F$_{unabs}$} & \colhead{logL$_{X}^{unabs}$} &\colhead{Classification}\\
\colhead{} & \colhead{} & \colhead{($\times$ 10$^{22}$ $cm^{-2}$)} & \colhead{(Index/keV)}  & \colhead{(10$^{-5}$ A)}  &\colhead{or Cstat}  &\colhead{(for Cstat)} & \multicolumn{2}{c}{($\times 10^{-15}$ erg s$^{-1} cm^{-1}$)} & \colhead{(erg s$^{-1}$)} &\colhead{}}
\startdata
\cutinhead{chi-square statistics} 
LZB$\_$24 (2030)   &PL+apec    &$0.28^{+0.30}_{-0.10}$     &$\Gamma=1.69^{+0.41}_{-0.18}$&$1.30^{+0.76}_{-0.48}$  &31.07/23     &\nodata &72      &130      &38.5 & XRB \tablenotemark{c}\\
\nodata        &\nodata    &\nodata                    &kT=$0.18^{+0.08}_{-0.05}$        &$2.50^{+58.6}_{-2.34}$  &\nodata      &\nodata &\nodata &\nodata  &\nodata &\nodata\\
(4793)         &\nodata    &\nodata                    &\nodata                          &0.575                   &\nodata      &\nodata &42      &75       &38.3 &\nodata\\
(5197)         &\nodata    &\nodata                    &\nodata                          &0.488                   &\nodata      &\nodata &35      &64       &38.2 &\nodata\\
LZB$\_$25 (2030)   &PL         &$0.12^{+0.02}_{-0.02}$   &$1.68^{+0.13}_{-0.12}$         &$4.90^{+0.68}_{-0.59}$  &105.5/66     &\nodata &290     &340      &38.9 & XRB\\
(4793)         &\nodata    &\nodata                    &\nodata                          &0.192                   &\nodata      &\nodata &55      &66       &38.2 &\nodata\\
(5197)         &\nodata    &\nodata                    &\nodata                          &0.904                   &\nodata      &\nodata &260     &310      &38.9 &\nodata\\
LZB$\_$26 (2030)   &PL         &$0.27^{+0.07}_{-0.09}$   &$2.15^{+0.50}_{-0.40}$         &$0.84^{+0.53}_{-0.30}$  &12.4/11     &\nodata  &43.8    & 25.9    &37.8 & XRB \tablenotemark{e}\\
\enddata
\tablecomments{Description of columns: Column 1: Source ID. The numbers in parenthesis show the observation ID used, Column 2: Fitted model, Column 3: H{\sc i} column Density, Column 4: Best fit value of the power law photon index or the thermal component temperature in keV, Column 5: Model normalization for the first observation used. In the case of multiple observations the number presented in column 5 is the conversion factor which is multiplied to the model normalization. The latter is given in units of $10^{-5}$ K, where K is for apec: $10^{-14} (4 \pi D_{A}^2)^{-1} \int n_{e} n_{p}$dV; $D_{A}$ is the angular distance to the source in centimeters and $n_{e} n_{p}$ are the electron and proton densities respectively in $cm^{-2}$, for PL: photons keV$^{-1}$ cm$^{-2}$ at 1 keV and for bbody: 10$^{39}$ erg s$^{-1}$ (10 kpc)$^{-1}$, Column 6: $\chi^2$/degrees of freedom or Cash statistic, Column 7: Goodness-of fit for Cash statistics, Column 8: Absorbed flux in 0.3 - 10 keV band, Column 9: Unabsorbed flux in 0.3 - 10 keV band, Column 10: Source classification}
\tablenotetext{a}{Radio candidate SNR by \citet{Vukotic05}}
\tablenotetext{b}{Radio candidate SNR by \citet{CW09}}
\tablenotetext{c}{Radio classified H{\sc ii} regions by \citet{CW09}}
\tablenotetext{d}{Fixed to Galactic line-of-sight value}
\tablenotetext{e}{Radio SNR/H{\sc ii} detected by \citet{CW09}}
\end{deluxetable}

\begin{deluxetable}{rcccccccccc}
\rotate
\tablecolumns {11}
\tabletypesize{\scriptsize}
\tablecaption{Spectral fitting results of the SNR candidates in NGC 2403}
\tablewidth{0pt}
\tablehead{\colhead{Source} & \colhead{Model} & \colhead{N$_{\rm H}$} & \colhead{$\Gamma$/kT} & \colhead{Norm} & \colhead{$\chi^2$/dof} & \colhead{Goodness($\%$)} &\colhead{F$_{abs}$}  &\colhead{F$_{unabs}$} & \colhead{logL$_{X}^{unabs}$} &\colhead{Classification}\\
\colhead{} & \colhead{} & \colhead{($\times$ 10$^{22}$ $cm^{-2}$)} & \colhead{(Index/keV)}  & \colhead{($10^{-5}$ A)}  &\colhead{or Cstat}  &\colhead{(for Cstat)} &\multicolumn{2}{c}{($\times 10^{-15}$erg s$^{-1} cm^{-1}$)} & \colhead{(erg s$^{-1}$)} &\colhead{}}
\startdata
\cutinhead{chi-square statistics} 
LZB$\_$21 (4630) &PL             &$0.14^{+0.07}_{-0.08}$   &$2.06^{+0.20}_{-0.14}$   &$1.02^{+0.40}_{-0.28}$   &8.3/14   &\nodata  &40  &55   &37.8  &XRB\\
LZB$\_$54 (4628) &PL             &$0.25^{+0.22}_{-0.15}$  &$1.37^{+0.32}_{-0.40}$   &$0.23^{+0.15}_{-0.08}$    &16.2/11  &\nodata  &18  &22   &37.4  &XRB\\
(4629)           &PL             &\nodata                  &\nodata                  &1.16                     &\nodata  &\nodata  &21  &25   &37.5  &\nodata\\
(4630)           &PL             &\nodata                  &\nodata                  &1.50                     &\nodata  &\nodata  &28  &33   &37.6  &\nodata\\
LZB$\_$58 (4630) &PL             &$0.12^{+0.31}_{-0.12}$   &$1.32^{+0.54}_{-0.40}$   &$0.38^{+0.35}_{-0.14}$   &4.77/5   &\nodata  &35  &39   &37.7  &XRB\\ 
LZB$\_$64 (2014) &PL + apec      &0.0436\tablenotemark{d}  &$\Gamma=1.95^{+0.23}_{-0.26}$    &$0.56^{+0.09}_{-0.12}$   &40.99/33   &\nodata   &34      &39      &37.7  &XRB\\
\nodata          &\nodata        &\nodata                  &kT=$0.82^{+0.16}_{-0.11}$          &$0.26^{+0.09}_{-0.09}$   &\nodata  &\nodata   &\nodata &\nodata &\nodata &\nodata\\
(4630)           &\nodata        &\nodata                  &\nodata                  &2.06                     &\nodata  &\nodata  &71  &80   &38.0  &\nodata\\
\enddata
\tablecomments{For the description of the Columns see Table 20}
\tablenotetext{a}{Optically detected SNRs by \citet{MFBL97}}
\tablenotetext{b}{Radio detected SNR by \citet{TH94}}
\tablenotetext{c}{Radio detected SNR by \citet{Eck02}}
\end{deluxetable}

\begin{deluxetable}{rcccccccccc}
\rotate
\tablecolumns {11}
\tabletypesize{\scriptsize}
\tablecaption{Spectral fitting results of the point-sources in NGC 5204}
\tablewidth{0pt}
\tablehead{
\colhead{Source} & \colhead{Model} & \colhead{N$_{\rm H}$} & \colhead{$\Gamma$/kT} & \colhead{Norm} & \colhead{$\chi^2$/dof} & \colhead{Goodness$\%$} &\colhead{F$_{abs}$}  &\colhead{F$_{unabs}$} & \colhead{logL$_{X}^{unabs}$} &\colhead{Classification}\\
\colhead{} & \colhead{} & \colhead{($\times$ 10$^{22}$ $cm^{-2}$)} & \colhead{(Index/keV)}  & \colhead{($10^{-5}$ A)}  &\colhead{or Cstat}  &\colhead{(for Cstat)} &\multicolumn{2}{c}{($\times 10^{-15}$erg s$^{-1} cm^{-1}$)} & \colhead{(erg s$^{-1}$)} &\colhead{}} 
\startdata
\cutinhead{chi-square statistics} 
LZB$\_$10 (2028) &PL          &$0.16^{+0.01}_{-0.01}$     &$2.8^{+0.05}_{-0.04}$         &$79.7^{+3.02}_{-3.04}$  &1635.6/1069&\nodata   &1839    &3937    &40.0 &ULX, XRB\\ 
(2029)       &\nodata         &\nodata                    &\nodata                       & 0.38                   &\nodata    &\nodata   &696.8   & 1492   &39.6 &\nodata\\
(3933)       &\nodata         &\nodata                    &\nodata                       & 0.58                   &\nodata    &\nodata   &1064    & 2278   &39.8 &\nodata\\
(3934)       &\nodata         &\nodata                    &\nodata                       & 0.28                   &\nodata    &\nodata   &521.1   & 1116   &39.5 &\nodata\\
(3935)       &\nodata         &\nodata                    &\nodata                       & 0.35                   &\nodata    &\nodata   &651.1   & 1394   &39.6 &\nodata\\
(3936)       &\nodata         &\nodata                    &\nodata                       & 1.29                   &\nodata    &\nodata   &2382    &5101    &40.2 &\nodata\\
(3937)       &\nodata         &\nodata                    &\nodata                       & 1.24                   &\nodata    &\nodata   &2286    & 4894   &40.1 &\nodata\\
(3938)       &\nodata         &\nodata                    &\nodata                       & 0.28                   &\nodata    &\nodata   &507     & 1085   &39.5 &\nodata\\
\enddata
\tablecomments{For the description of the Columns see Table 20}
\end{deluxetable}

\clearpage

\begin{deluxetable}{ccccc}
\tablecolumns {5}
\tabletypesize{\scriptsize}
\tablecaption{Composite SNR/XRB sources}
\tablewidth{0pt}
\tablehead{
\colhead{Galaxy} & \colhead{Source ID}& \colhead{apec luminosity} & \colhead{PL luminosity } & \colhead{Total luminosity}\\
\colhead{} & \colhead{} & \colhead{($\times$10$^{37}$ erg s$^{-1}$)} & \colhead{($\times$10$^{37}$ erg s$^{-1}$)} & \colhead{($\times$10$^{37}$ erg s$^{-1}$)}}
\startdata
NGC 2403 & 93  & $-$            & 2.4 (100$\%$)     &2.4 \\
NGC 2403 & 99  & 2.6 (14$\%$) & 16.1 (86$\%$)     &18.7 \\
NGC 2403 & 104 & $-$            & 92.1 (100$\%$)    &92.1 \\ 
NGC 4214 & 26  & $-$            & 10.4 (100$\%$)    &10.4  \\
\enddata
\tablenotetext{\ast}{The numbers in parenthesis indicate the contribution of the luminosity of each component to the total absorption--corrected luminosity. The three X-ray SNRs in NGC 2403 are optically known SNRs while the one in NGC 4214 is associated with a radio SNR/H{\sc ii} source.}
\end{deluxetable}

\begin{deluxetable}{ccccc}
\tablecolumns {5}
\tabletypesize{\scriptsize}
\tablecaption{Number of the SNRs in the MCs and M33}
\tablewidth{0pt}
\tablehead{
\colhead{Galaxy} & \colhead{L$_{FIR}$} & \colhead{Observed SNRs} & \colhead{Rescaled Number of SNRs} & \colhead{Rescaled Number of SNRs}\\
\colhead{} & \colhead{($\times10^{42}$ erg s$^{-1}$)} & \colhead{} & \colhead{($\alpha= -0.5$)} & \colhead{($\alpha= -1$)}}
\startdata
\cutinhead{Luminosity distribution of \citet{Ghavamian05}}
LMC &1.96 & 31.6 &4.5 &\nodata \\
SMC &0.29 & 8.9 &1.3 &\nodata \\
\cutinhead{Luminosity distribution of \citet{HaP01}}
M33 & 2.86  & 9.1 & 2.6  & 0.8  \\
\enddata
\tablecomments{The LMC, SMC and M33 FIR luminosities correspond to the 42-122 $\mu$m band and are adopted from \citet{Rice88}. The numbers of the observed SNRs are down to the detection limit of each respective study: for LMC--SMC is 10$^{35}$ erg s$^{-1}$ and for M33 is 8.5$\times$10$^{35}$ erg s$^{-1}$. The observed numbers of the MC SNRs are rescaled to the numbers of SNRs down to our limiting luminosity of 5$\times$10$^{36}$ erg s$^{-1}$ while the observed M33 SNRs are rescaled to the limiting luminosity of \citealt{P07} (10$^{37}$ erg s$^{-1}$).}
\end{deluxetable}

\begin{deluxetable}{lcccc}
\tablecolumns {5}
\tabletypesize{\scriptsize}
\tablecaption{Expected number of SNRs (assuming an MC XLF)}
\tablewidth{0pt}
\tablehead{
\colhead{Galaxy} & \colhead{L$_{FIR}$} & \colhead{(\case{SNR}{L_{FIR}})$_{MC}$} & \colhead{Expected Number of SNRs}  & \colhead{Observed SNRs}\\
& & & &\\
&\colhead{($\times10^{42}$ erg s$^{-1}$)} &\colhead{($\times10^{-42}$ SNR / (erg s $^{-1}$))} & &\\
}
\startdata
NGC 3077 &0.94 &3.38 &3.2   & 5\\
NGC 4449 &4.10 &3.38 &13.8  & 3\\
NGC 4395 &0.24 &3.38 &0.8   & 1\\
NGC 4214 &2.52 &3.38 &8.5   & 5\\
NGC 5204 &0.39 &3.38 &1.3   & 0\\
NGC 2403 &4.33 &3.38 &14.6  & 8\\
\enddata
\tablecomments{Column 1: Galaxy ID, Column 2: The 42-122 $\mu$m FIR luminosities of the galaxies based on integrated flux measurements with IRAS in the 60 and 100 $\mu$m bands \citep{Ho97}, using the calibration of \citet{Rice88} and the distances in Table 1. Column 3: The average number of MC SNRs (down to 5$\times$10$^{36}$ erg s$^{-1}$) per unit FIR luminosity; see $\S$5.3.2 for details, Column 4: The expected number of SNRs for the galaxies in our sample, derived by rescaling the number of the MC SNRs, Column 5: The observed SNRs for each galaxy down to our detection limit of 5$\times$10$^{36}$ erg s$^{-1}$.}
\end{deluxetable}

\begin{deluxetable}{lcccccc}
\rotate
\tablecolumns {7}
\tabletypesize{\scriptsize}
\tablecaption{Expected number of SNRs in spiral galaxies (assuming an M33-like XLF) }
\tablewidth{0pt}
\tablehead{
\colhead{Galaxy} & \colhead{L$_{FIR}$} & \colhead{(\case{SNR}{L_{FIR}})$_{M33}$}  & \colhead{Expected Number of SNRs} & \colhead{(\case{SNR}{L_{FIR}})$_{M33}$} &\colhead{Expected Number of SNRs} & \colhead{Observed SNRs}\\
& & & & \\
&\colhead{($\times10^{42}$ erg s$^{-1}$)} &\colhead{($\times10^{-42}$ SNR / (erg s$^{-1}$))} & &\colhead{($\times10^{-42}$ SNR / (erg s$^{-1}$))}\\
& & & & \\
& &\colhead{($\alpha$ = - 0.5)} &\colhead{($\alpha$ = - 0.5)} &\colhead{($\alpha$ = - 1)} &\colhead{($\alpha$ = - 1)}
}
\startdata
NGC 2403 &4.33 &0.93  &4.0  &0.27  &1.2  &3\\
M 101    &37.2 &0.93  &34.6 &0.27  &10.0 &7\\
NGC 6946 &36.0 &0.93  &33.5 &0.27  &9.7  &6\\
NGC 4736 &8.20 &0.93  &7.6  &0.27  &2.2  &5\\
\enddata
\tablecomments{Column 1: Galaxy ID, Column 2: The 42-122 $\mu$m FIR luminosities of the galaxies are based on integrated flux measurements with IRAS in the 60 and 100 $\mu$m bands \citep{Ho97} using the calibration of \citet{Rice88} and the distances from \citet{P07}, Column 3: Number of M33 SNRs per unit FIR luminosity, down to a limiting luminosity of 10$^{37}$ erg s$^{-1}$ \citep{P07} and assuming a cumulative slope of - 0.5, Column 4: The expected number of SNRs for $\alpha$ = -0.5, estimated by rescaling the M33 XLF by the FIR luminosity of each galaxy (see $\S$5.4 for details), Column 5: Number of M33 SNRs per unit FIR luminosity, down to a limiting luminosity of 10$^{37}$ erg s$^{-1}$ (\citealp{P07}) and assuming a cumulative slope of - 1, Column 6: Same as Column 4 but for a cumulative slope of $\alpha$ = -1, Column 7: Observed SNRs from the sample of galaxies of \citet{P07} (down to 10$^{37}$ erg s$^{-1}$).}
\end{deluxetable}

\newpage

\begin{figure*}
\epsscale{2.0}
\plotone{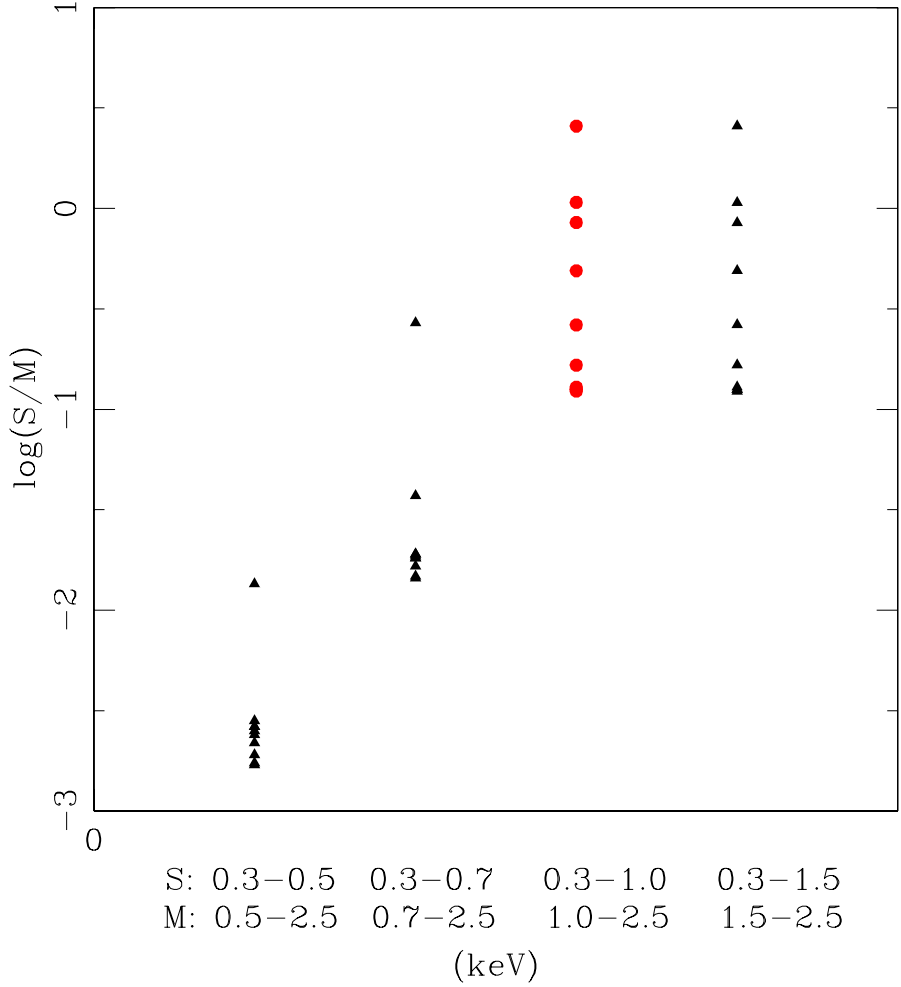}
\caption{Soft/Medium color distribution for different choices of the soft and medium bands. The abscisa shows the different selections for the two bands, while different points show the S/M color which correspond to different choices of temperature (0.25 - 2.0 keV from the bottom towards the top of each data set). The maximum discrimination of the S/M color distribution is seen for the last two energy band choices (S$_3$ - M$_3$ and S$_4$ - M$_4$). These energy band selections can allow us distinguish emission lines coming from low temperature thermal emitting gas or harder emission from hotter thermal continuum. The circles denote the S/M color distribution of the two bands that were used by this study (S: 0.3 - 1.0 keV and M: 1.0 - 2.5 keV).}
\end{figure*}

\begin{figure*}
\includegraphics[angle=270,scale=0.70]{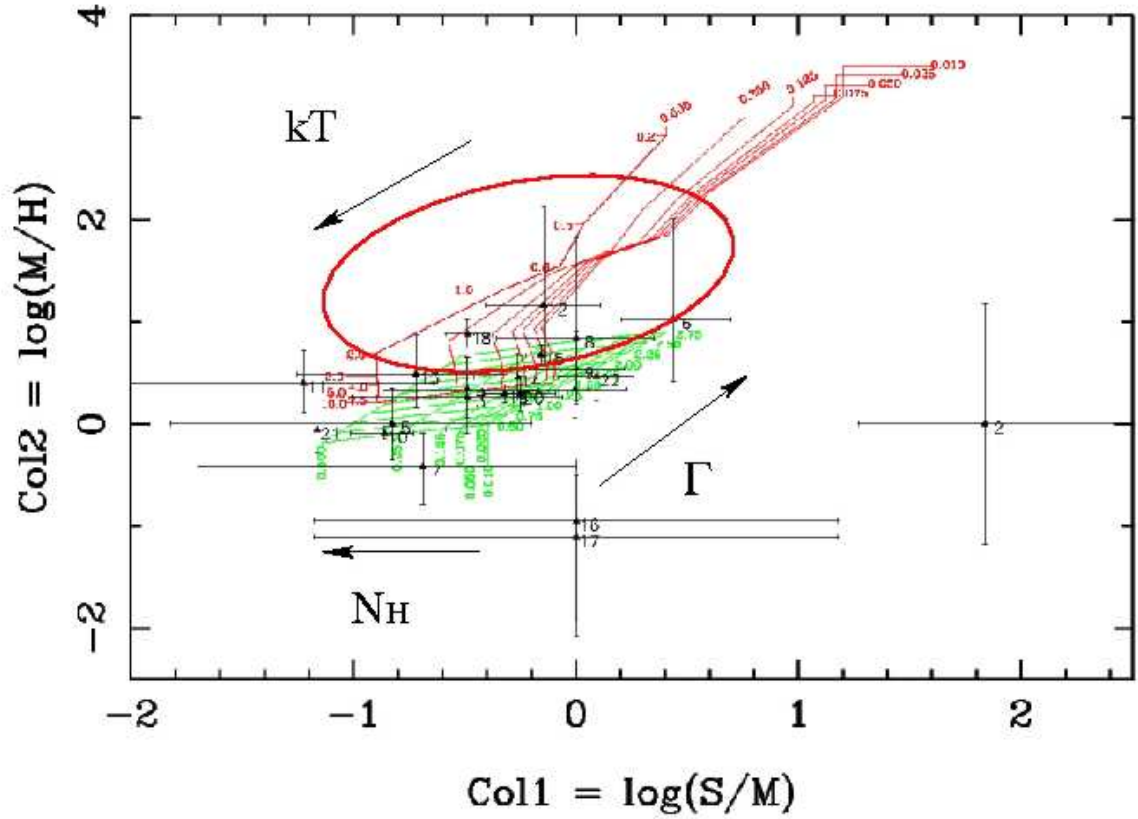}
\caption{Color--color diagram of the detected X-ray sources in NGC 3077. We overplot grids for thermal (red) and power--law (green) models for different values of temperature (kT), absorbing H{\sc i} column density (N$_{\rm H}$) and photon index $\Gamma$, calculated for the effective area at the reference point (see \S3.3). Sources that appear to have temperatures below 2 keV and mainly lie on the thermal grid (red 
locus of SNRs) are potential SNRs.[See the elctronic edition for a color version of this figure.]}
\end{figure*}

\begin{figure*}
\includegraphics[angle=270,scale=0.65]{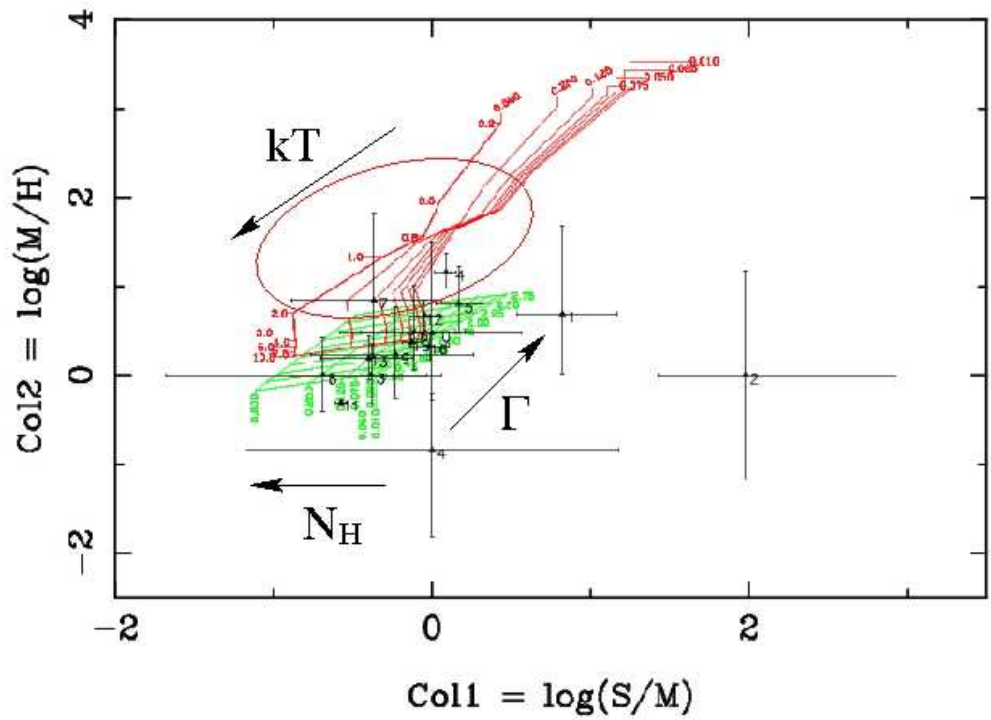}
\caption{Same as Fig. 2 for NGC 4395.} 
\end{figure*}

\begin{figure*}
\includegraphics[angle=270,scale=0.60]{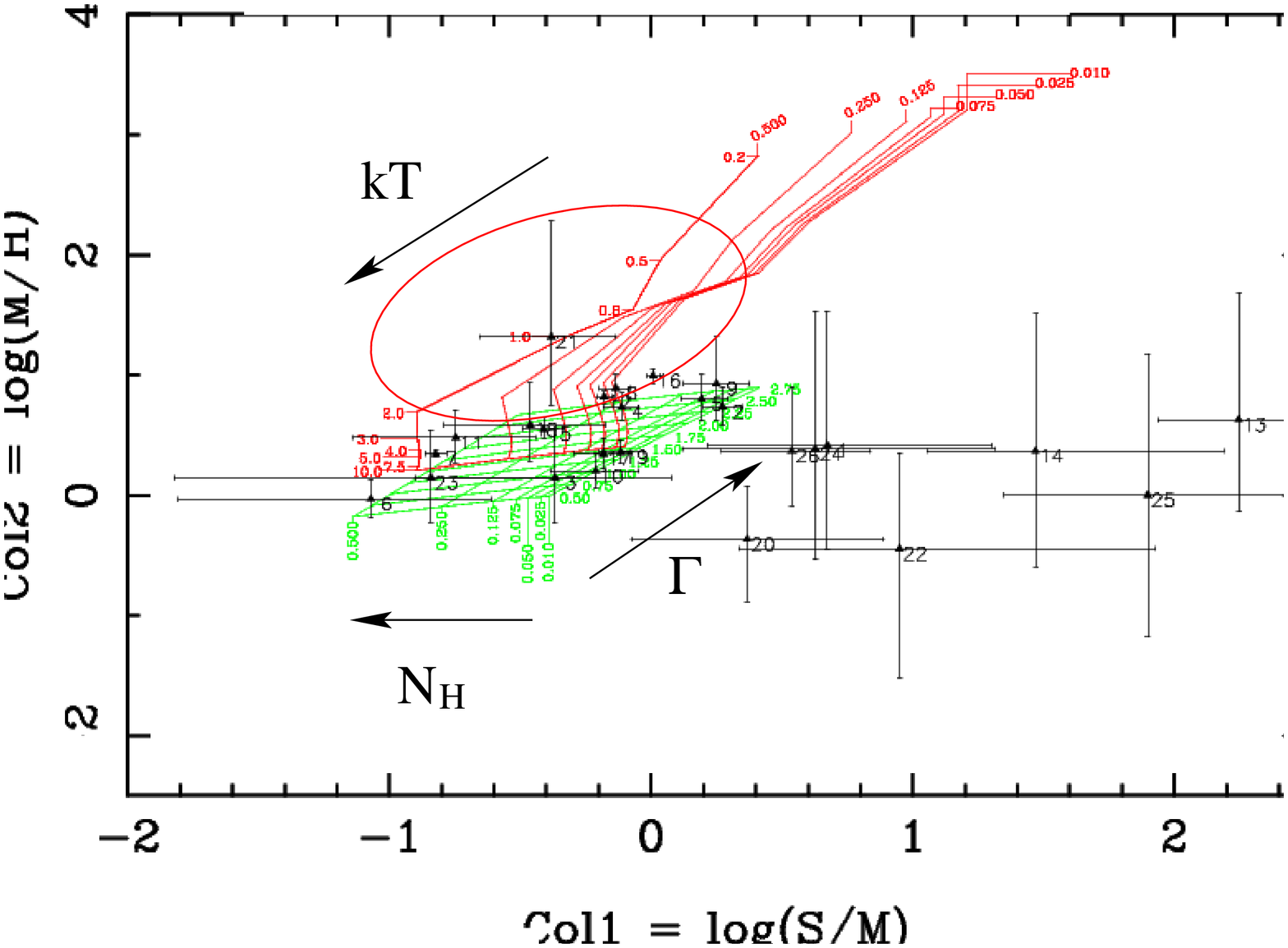}
\caption{Same as Fig. 2 for NGC 4449}
\end{figure*}

\begin{figure*}
\includegraphics[angle=270,scale=0.65]{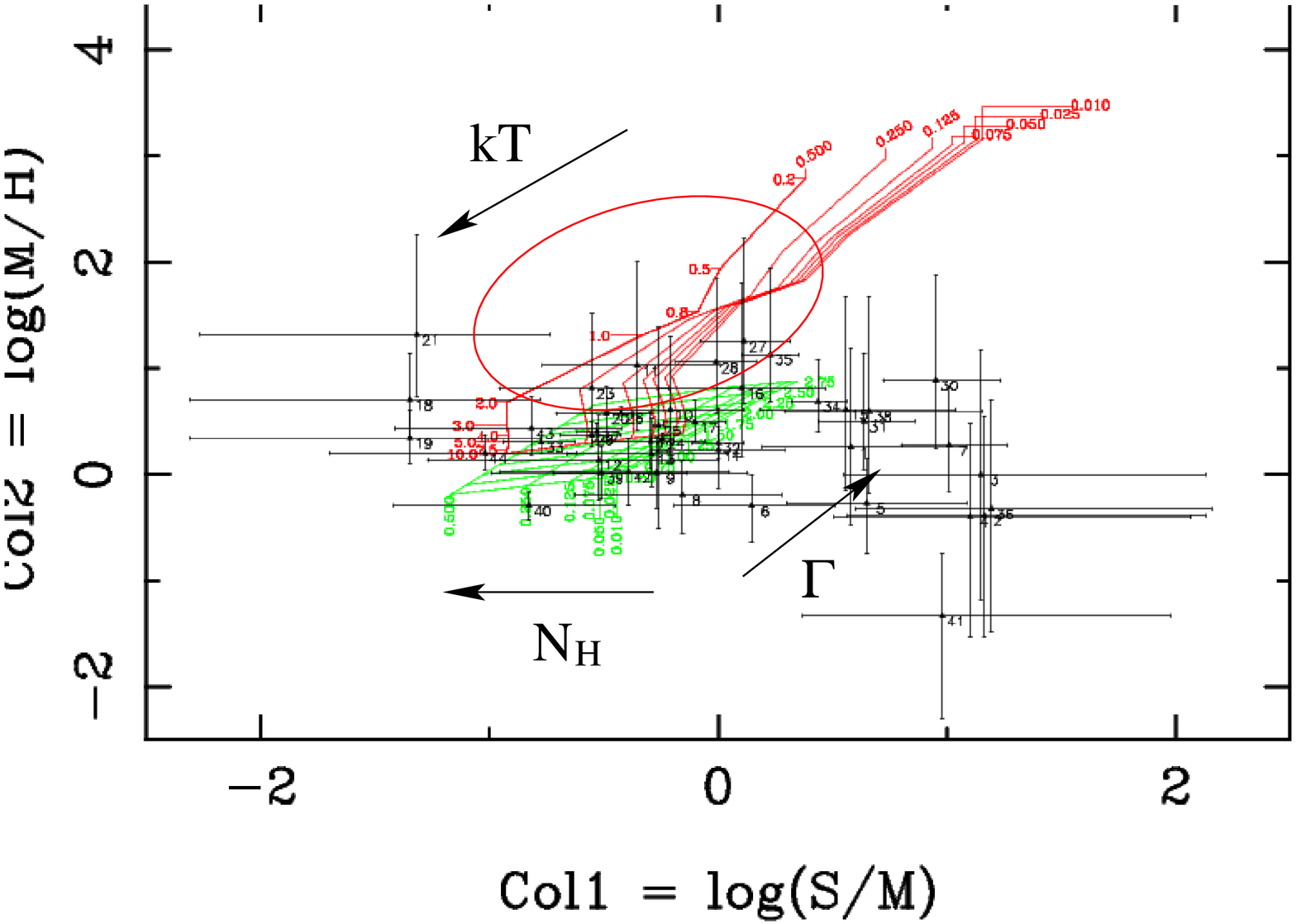}
\caption{Same as Fig. 2 for NGC 4214} 
\end{figure*}

\begin{figure*}
\includegraphics[angle=270,scale=0.65]{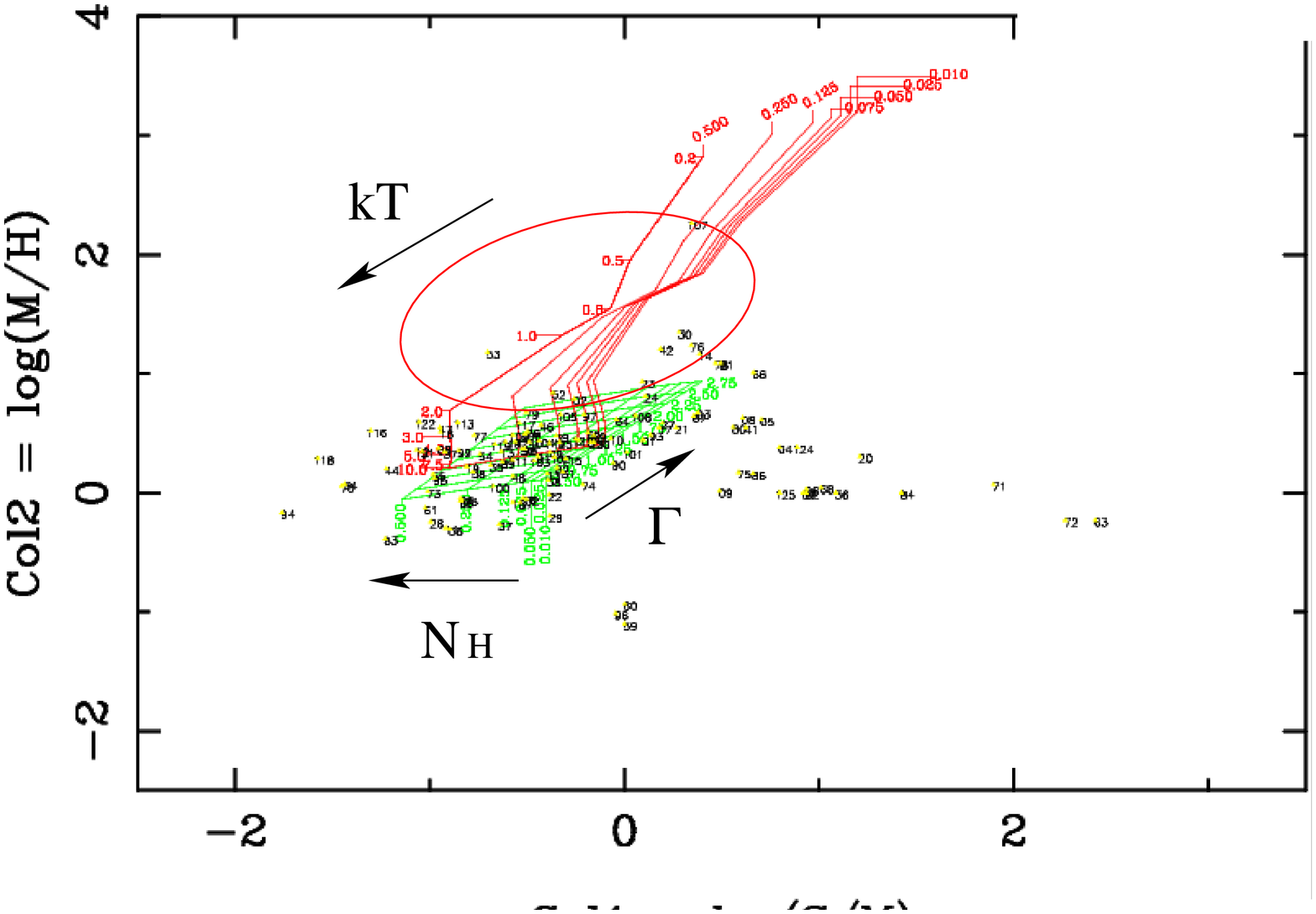}
\caption{Same as Fig. 2 for NGC 2403. Errors were not displayed for clarity.} 
\end{figure*}

\begin{figure*}
\includegraphics[angle=270,scale=0.65]{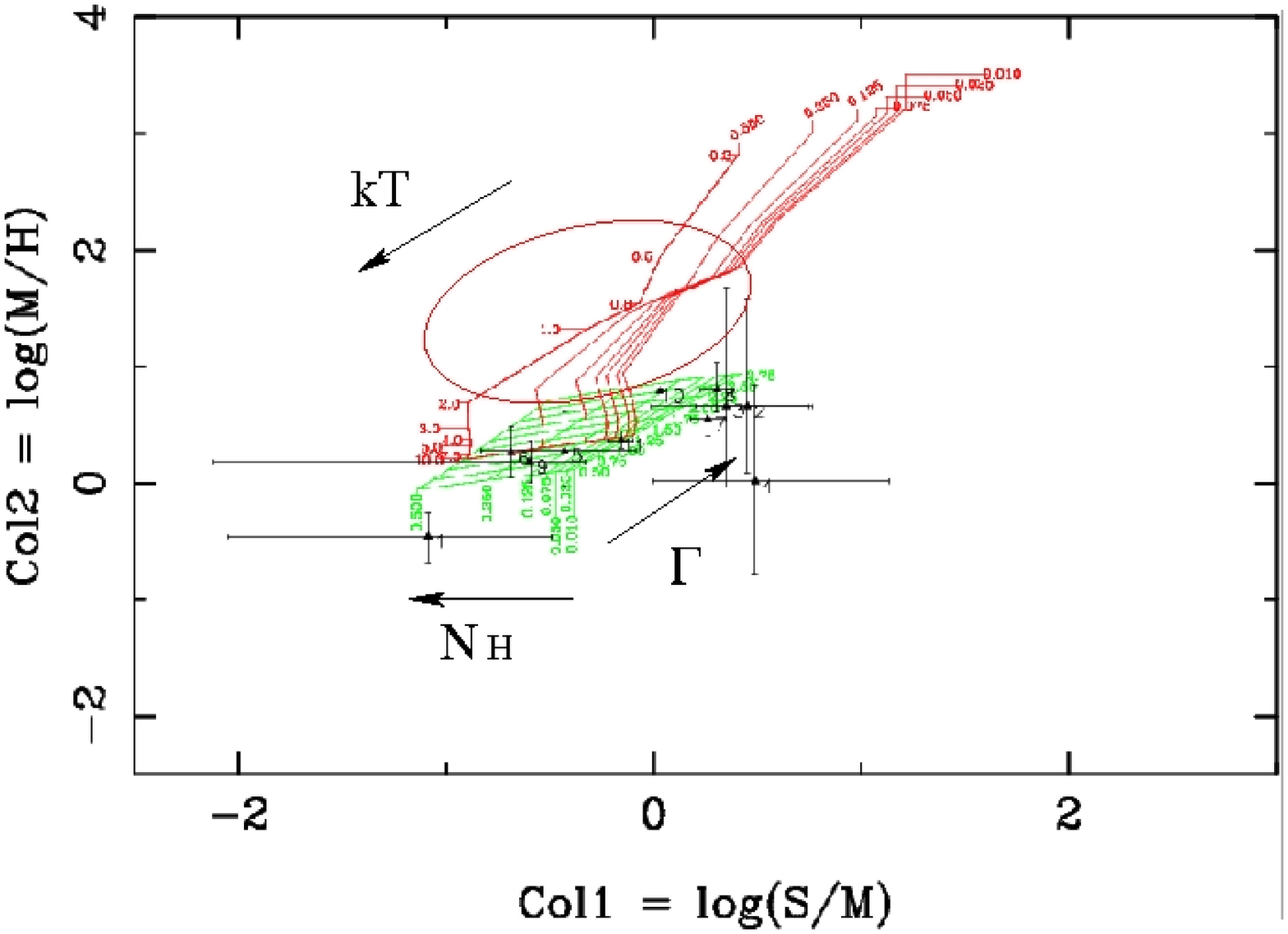}
\caption{Same as Fig. 2 for NGC 5204} 
\end{figure*}

\begin{figure*}
\epsscale{2.0}
\plotone{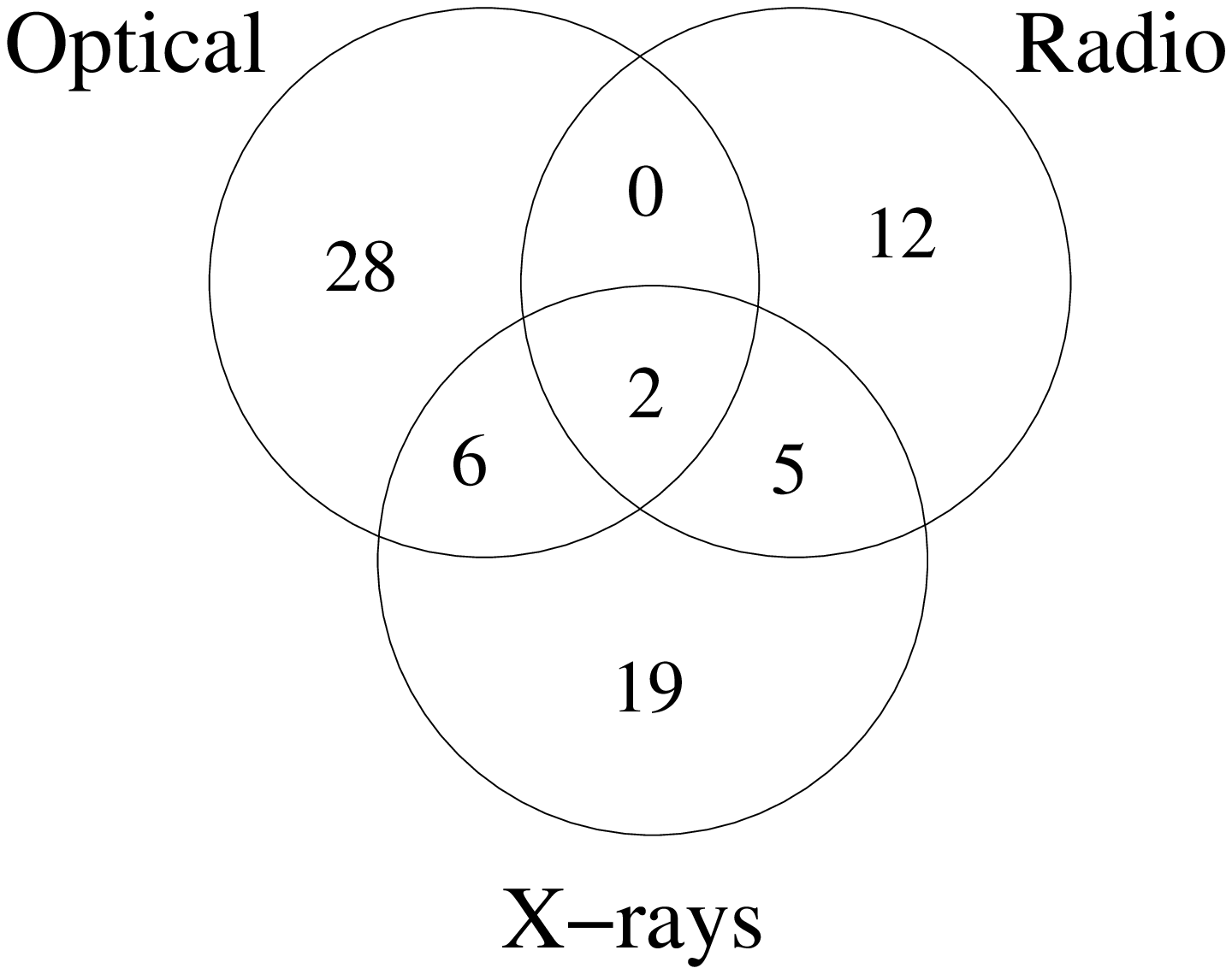}
\caption{Venn diagram for all SNRs detected in our sample of galaxies}
\end{figure*}

\begin{figure*}
\epsscale{2.0}
\plotone{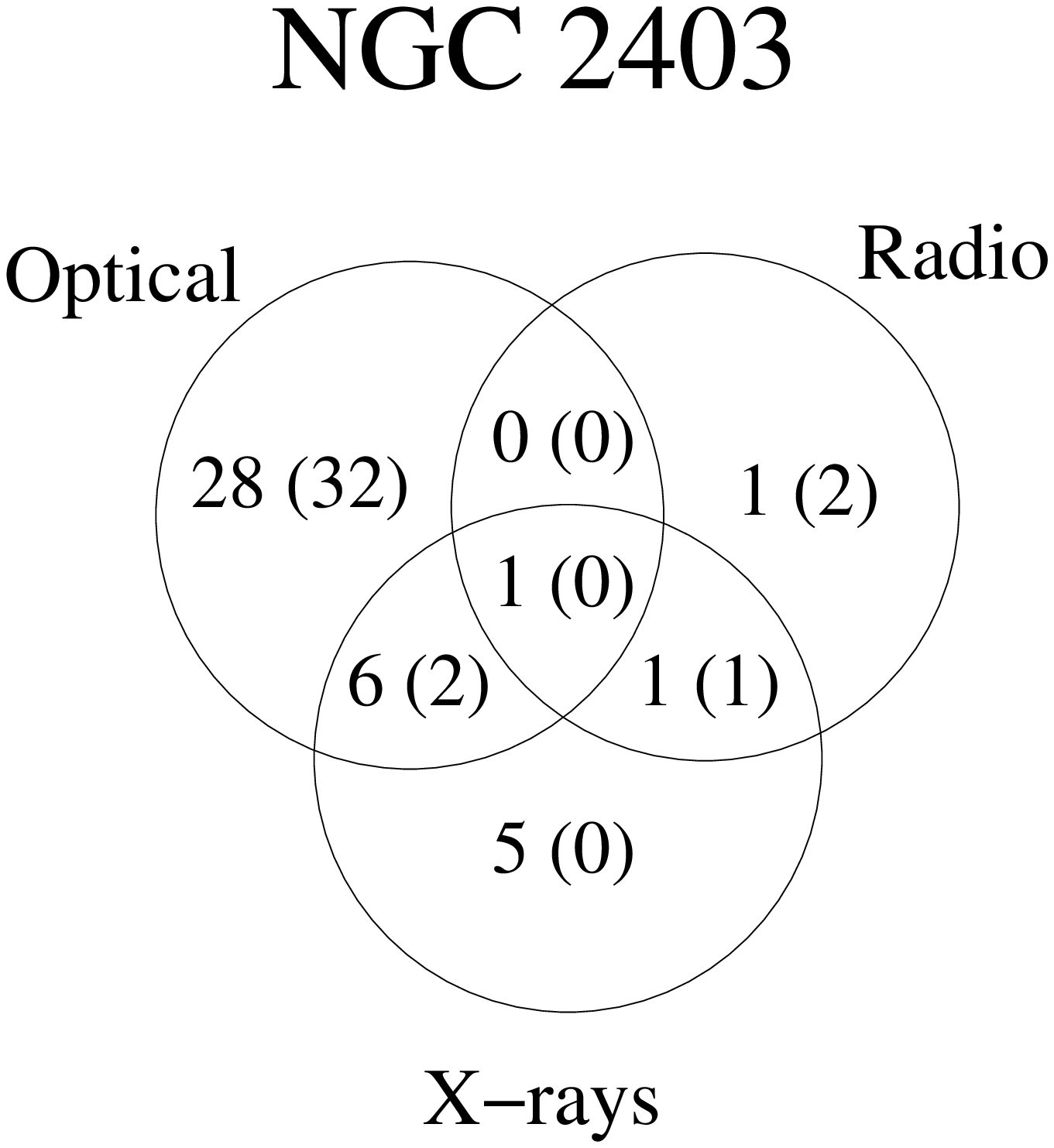}
\caption{Venn diagram for NGC 2403. The numbers in parenthesis refer to the optically/radio SNRs from \citet{P07}.}
\end{figure*}

\clearpage

\begin{figure*}
\epsscale{2.0}
\plotone{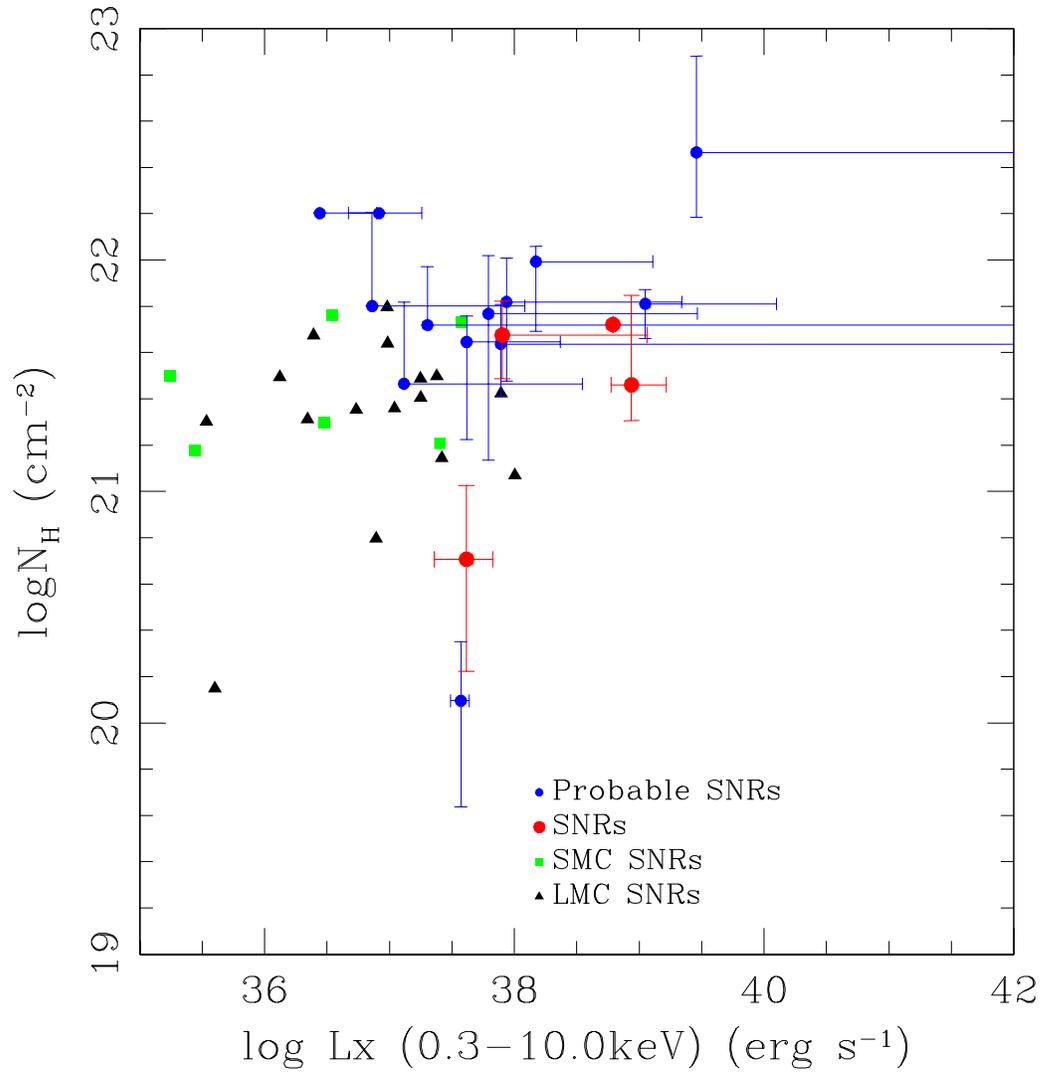}
\caption{The H{\sc i} column density (corrected for the Galactic column) plotted against the absorption--corrected luminosity. Squares and triangles denote the X-ray SNRs for SMC and LMC respectively while circles represent the X-ray SNRs of this study.}
\end{figure*}       	   

\begin{figure*}
\epsscale{2.0}
\plotone{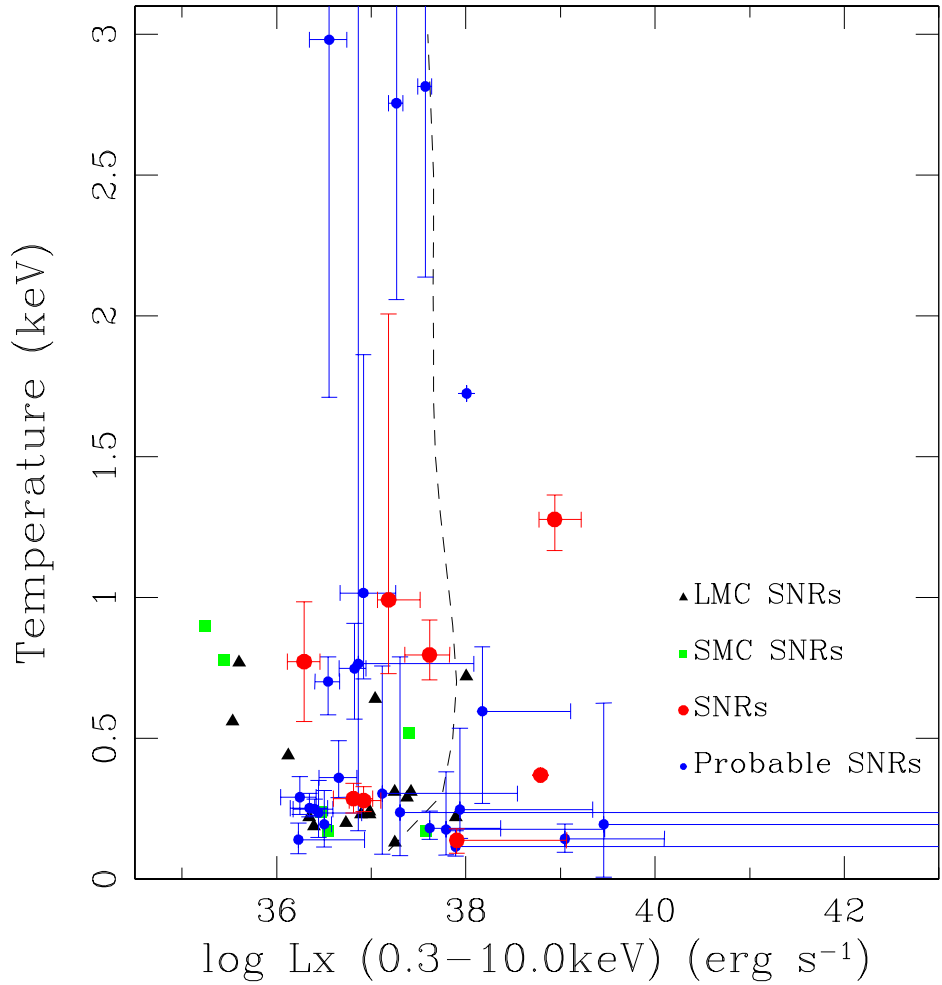}
\caption{Temperature of the X--ray selected SNRs in this study against their absorption--corrected X--ray luminosity. Circles are for the SNRs of this study, squares for the SMC SNRs and triangles for the LMC SNRs. The dashed line shows the expected relation between temperature and luminosity for a thermal source at a distance of 5 Mpc, based on an apec model with a fixed emission measure (EM).} 
\end{figure*}

\begin{figure*}
\epsscale{2.0}
\plotone{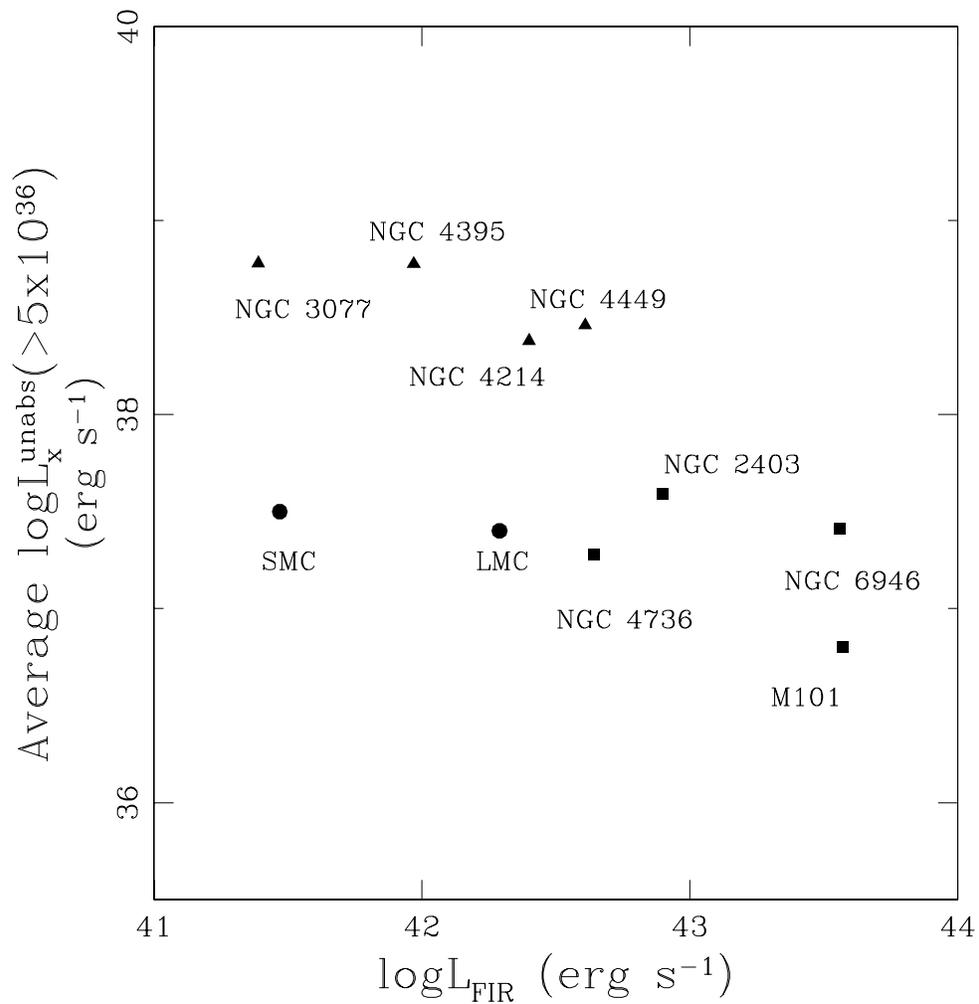}
\caption{The average absorption--corrected X--ray luminosity in the 0.3 - 10 keV band for the SNRs in our sample. We only include objects above the completeness limit of this study (5$\times$10$^{36}$ erg s$^{-1}$). Squares and triangles represent spiral and irregular galaxies respectively while circles correspond to the Maggelanic Clouds.} 
\end{figure*}

\begin{figure*}
\epsscale{2.0}
\plotone{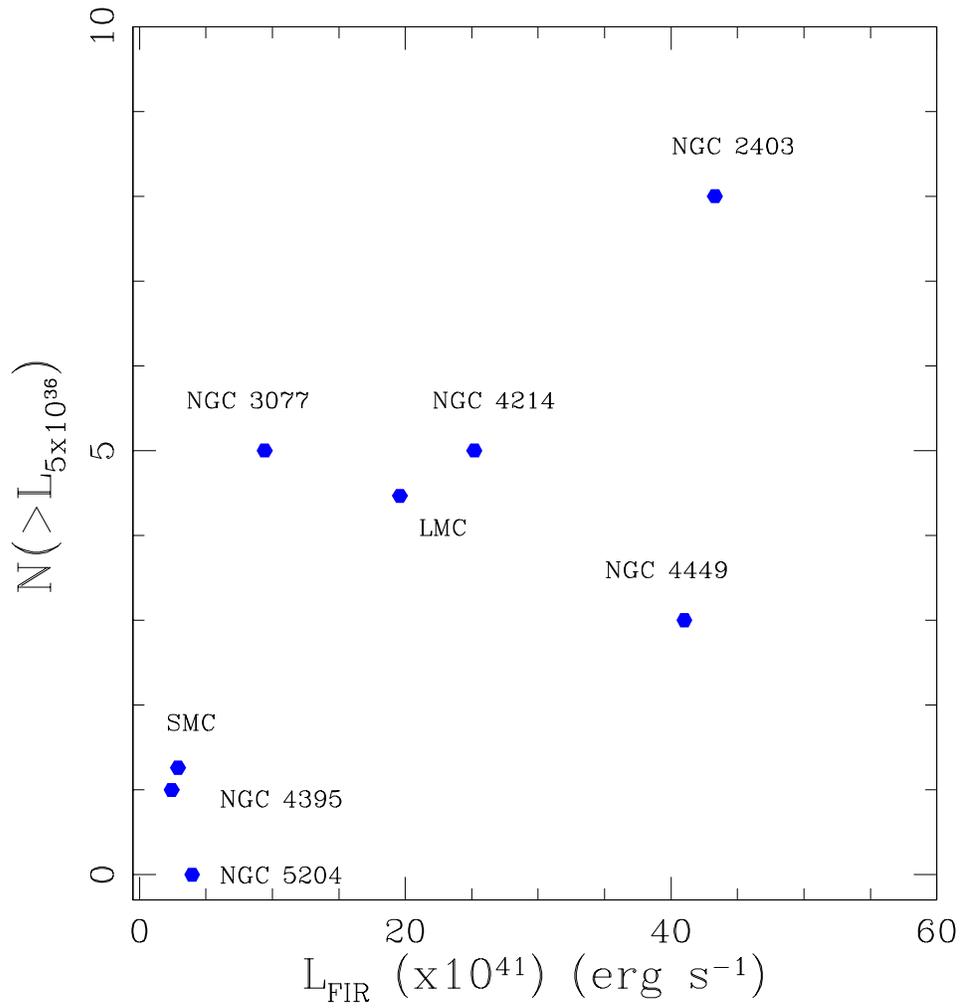}
\caption{Number of SNRs above the completeness limit of this study against the integrated Far Infrared (42--122$\mu$m) luminosity (see text for details).} 
\end{figure*}

\begin{figure*}
\epsscale{2.0}
\plotone{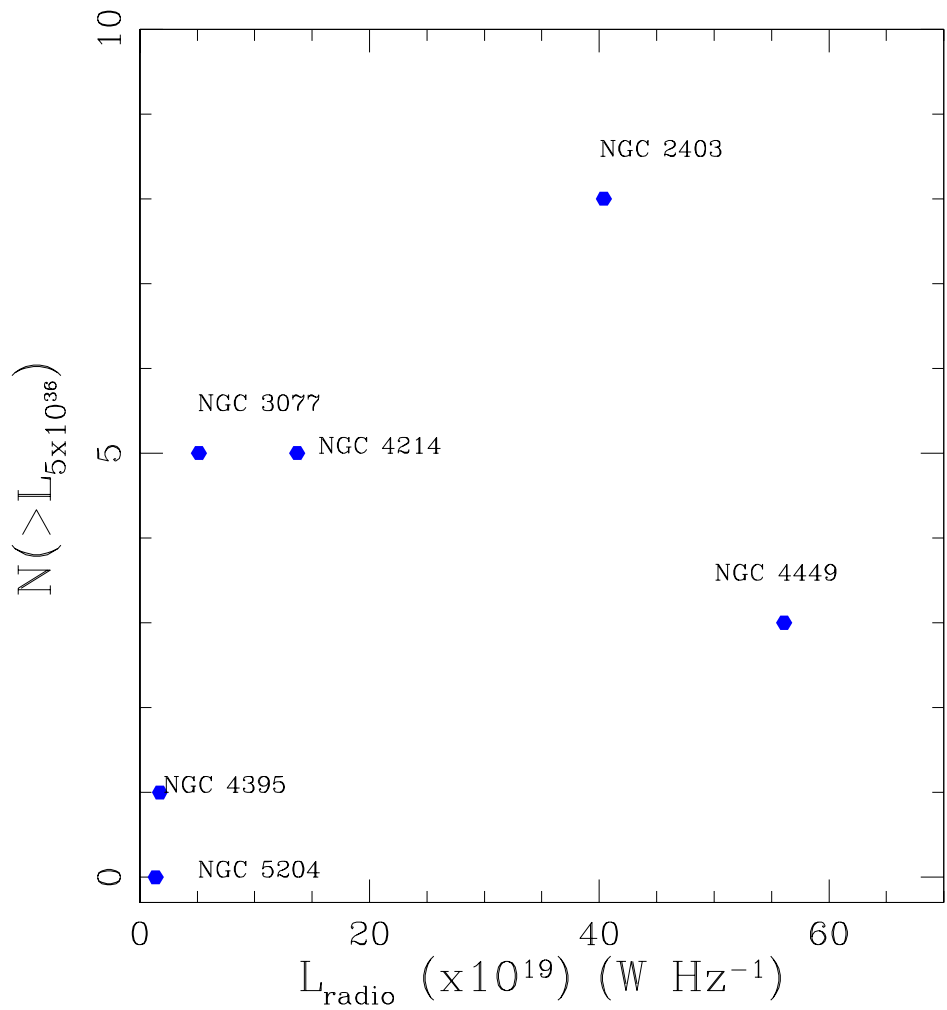}
\caption{Number of SNRs above the completeness limit of this study against the radio luminosity.} 
\end{figure*}

\end{document}